\documentclass[a4paper,11pt]{article}
\pdfoutput=1 
\usepackage{jcappub}
\usepackage{amsmath}
\usepackage{mathrsfs} 
\usepackage{amsfonts}
\usepackage{amssymb}
\usepackage{amsthm}
\usepackage{mathtools}
\usepackage{graphicx}
\usepackage{latexsym}
\usepackage{xspace}
\usepackage{hyperref} 
\usepackage{bm}
\usepackage{relsize}
\usepackage{tabularx}
\usepackage{multirow}
\usepackage{amssymb}
\usepackage{lscape}
\usepackage[most]{tcolorbox}
\usepackage{comment}
\usepackage[utf8]{inputenc}
\usepackage[english]{babel}
\usepackage{slashed}
\usepackage{empheq}
\usepackage[makeroom]{cancel}
\newcolumntype{P}[1]{>{\centering\arraybackslash}p{#1}}
\usepackage{braket}

\newcommand{\bfk}{{\mathbf{k}}}

\newcommand{\e}{\epsilon}

\usepackage{floatrow}
\usepackage{graphicx}
\usepackage[label font=bf,labelformat=simple]{subfig}
\usepackage{upgreek}

\usepackage[margin=1in,includefoot]{geometry}
\usepackage{xfrac} 
\usepackage{tikz-feynman}
\usepackage{tikz}	
\tikzfeynmanset{compat=1.0.0}

\allowdisplaybreaks

\makeatletter
\gdef\@fpheader{}
\g@addto@macro\bfseries{\boldmath}
\makeatother

\makeatletter
\DeclareRobustCommand*\uell{{\mathpalette\@uell\relax}}
\newcommand*\@uell[2]{
\setbox0=\hbox{#1\ell#1\ell}
\setbox1=\hbox{\rotatebox{10}{#1\ell#1\ell}}
\dimen0=\wd0 \advance\dimen0 by -\wd1 \divide\dimen0 by 2
\mathord{\lower 0.1ex \hbox{\kern\dimen0\unhbox1\kern\dimen0}}
}

\def\phia{\varphi_{a}}
\def\phir{\varphi_{r}}
\def\dphir{\dot{\varphi}_{r}}
\def\dphia{\dot{\varphi}_{a}}

\renewcommand{\H}{\mathcal{H}}
\renewcommand{\O}{\mathcal{O}}

\newcommand{\then}{\quad \Rightarrow\quad}
\newcommand{\exx}[1]{\langle\!\langle #1 \rangle\!\rangle}
\newcommand{\T}{\mathcal{T}}

\usepackage{colortbl}
\definecolor{summersky}{cmyk}{0.71,0.33,0,0.5}
\definecolor{flamingo}{cmyk}{0,0.51,0.71,0.5}
\definecolor{rp}{cmyk}{0.2, 1, 0.6, 0}
\definecolor{pacificblue}{cmyk}{0.95,0.3,0, 0.5}
\definecolor{gray60}{cmyk}{0.4,0.4,0,0.8}

\newcommand{\ex}[1]{\langle #1 \rangle}

\renewcommand{\H}{\mathcal{H}}
\renewcommand{\O}{\mathcal{O}}

\newcommand{\pip}{\varphi_{+}}
\newcommand{\pim}{\varphi_{-}}

\newcommand{\pa}{\phi_{a}}
\newcommand{\pr}{\phi_{r}}
\newcommand{\Al}{A_{||}}
\newcommand{\At}{A_{\perp}}

\newcommand{\at}{a_{\perp}}

\newcommand{\fini}{f}

\newcommand\radot{%
  \mathord{%
    \mspace{1mu}%
    \text{\rado}%
    \mspace{1mu}%
  }%
}

\newcommand\rado{%
    \tikz[line cap=round,x=1ex,y=1ex,line width=0.4pt]
    {\draw (0,1) |- (1,0) (0.55,0) 
    ; \fill (0.23,0.23) 
    ;}%
}

\newcommand{\Tr}{\mathrm{Tr}}

\newcommand{\dd}{\mathrm{d}}
\newcommand{\ee}{e}

\newcommand{\bmx}{\boldmathsymbol{x}}

\newcommand{\Ima}{\Im \mathrm{m}\,}

\newcommand{\beq}{\begin{equation}}
	\newcommand{\eeq}{\end{equation}}
\newcommand{\bea}{\begin{equation}\begin{aligned}}
		\newcommand{\eea}{\end{aligned}\end{equation}}

\newlength{\wsingfig}
\newlength{\wdblefig}
\newlength{\wquadfig}
\newlength{\wtriplefig}
\newcommand{\Eq}[1]{Eq.~(\ref{#1})}
\newcommand{\Eqs}[1]{Eqs.~(\ref{#1})}
\newcommand{\Fig}[1]{Fig.~{\ref{#1}}}

\newcommand{\Sec}[1]{Sec.~\ref{#1}}

\usepackage{dsfont}

\def\bmx{{\boldsymbol{x}}}

\def\bmk{{\boldsymbol{k}}}

\subheader{}

\title{An Open Effective Field Theory for light in a medium}

\author[a]{Santiago Ag\"u\'i Salcedo,}
\author[a]{Thomas Colas,}
\author[a]{and Enrico Pajer}

\affiliation[a]{Department of Applied Mathematics and Theoretical Physics, University of Cambridge, Wilberforce Road, Cambridge, CB3 0WA, UK}
\emailAdd{sa2013@cam.ac.uk}
\emailAdd{tc683@cam.ac.uk}
\emailAdd{enrico.pajer@gmail.com}

\begin{document}
\sloppy

\abstract{In many scenarios of interest, a quantum system interacts with an unknown environment, necessitating the use of open quantum system methods to capture dissipative effects and environmental noise. With the long-term goal of developing a perturbative theory for open quantum gravity, we take an important step by studying Abelian gauge theories within the Schwinger-Keldysh formalism. We begin with a pedagogical review of general results for open free theories, setting the stage for our primary focus: constructing the most general open effective field theory for electromagnetism in a medium. We assume locality in time and space, but allow for an arbitrary finite number of derivatives. Crucially, we demonstrate that the two copies of the gauge group associated with the two branches of the Schwinger-Keldysh contour are not broken but are instead deformed by dissipative effects. We provide a thorough discussion of gauge fixing, define covariant gauges, and calculate the photon propagators, proving that they yield gauge-invariant results. A notable result is the discovery that gauge invariance is accompanied by non-trivial constraints on noise fluctuations. We derive these constraints through three independent methods, highlighting their fundamental significance for the consistent formulation of open quantum gauge theories.}

\maketitle


\section{Introduction}\label{sec:intro}


Effective Field Theories (EFTs) excel when our microscopic knowledge is incomplete, but fundamental physical principles constrain the dynamics sufficiently to parameterize our ignorance in a controlled and hierarchical way. In cosmology, some principles that are robust in flat spacetime rest on much shakier foundations. For instance, time translations and Lorentz boosts are spontaneously broken by the presence of a preferred reference frame in which the universe appears homogeneous and isotropic. Additionally, the lack of energy conservation in cosmology facilitates information loss between observable and inaccessible sectors (see e.g. \cite{Burgess:2024eng}), raising critical questions about the unitarity of effective cosmological descriptions. Open EFTs (see \cite{breuerTheoryOpenQuantum2002,kamenev_2011} and also \cite{Burgess:2022rdo, Colas:2024lse} for concise reviews) address these issues by constructing frameworks that incorporate dissipation and noise — features of the non-unitary evolution of the observed sector — in a systematic way. By leveraging unitarity of the UV complete theory, locality, and appropriate symmetries, these constructions provide a controlled approach to capturing these effects. This methodology has been successfully applied to the EFT of Inflation \cite{Salcedo:2024smn}, in the decoupling limit, where interactions with dynamical components of the metric can be safely ignored.

Extending this framework to late-time cosmology requires going beyond this limit and addressing the full complexity of dynamical gravity \cite{Gubitosi:2012hu}. A crucial preliminary step in this direction is ensuring a clear understanding of the theory's diffeomorphism covariance. An even more foundational step involves incorporating gauge symmetries into Open EFTs. Complementing the formal approaches developed in non-equilibrium EFTs and dissipative hydrodynamics \cite{Liu:2018kfw} — such as the in-in coset construction \cite{Akyuz:2023lsm} and Schwinger-Keldysh BRST quantization \cite{Haehl:2016pec} — this work aims to bridge formalism and practical implementation by providing an explicit and pedagogical example of such constructions.

In this article, we study the Abelian gauge theory of electromagnetism in a medium using the Schwinger-Keldysh formalism \cite{Schwinger:1960qe, Keldysh:1964ud}. Our goal is developing a bottom-up Open EFT for the propagation of light experiencing dissipation and noise. Assuming locality in time and space, we formulate the most general open effective functional consistent with the gauge invariance of physical observables. We analyze gauge transformations and gauge fixing in the presence of stochastic effects, reproducing well-established results from the literature \cite{Griffiths_2017}.

Our main findings are as follows. In the free theory, dissipative systems exhibit non-Hermitian Hessian matrices, which have profound implications for the structure of the path integral used to compute expectation values of observables in the presence of gauge symmetries. Notably, we identify the emergence of two distinct gauge groups, one acting on advanced and one on retarded fields. This is surprising because naively, the two copies of the gauge group associated to the two branches of the Schwinger-Keldysh path integral are explicitly broken to a single diagonal gauge group by dissipative effects. Here we show that, in addition to this unbroken diagonal gauge symmetry acting on retarded gauge fields, a new advanced gauge symmetry arises, which is deformed compared to the usual gauge transformations by the presence of dissipation. This structure provides sufficient freedom to independently gauge-fix the fields on both sides of the contour. Upon implementing this procedure, we recover the dispersion relations for the two helicities propagating through a medium that is possibly dispersive, dissipative, and anisotropic. Importantly, this framework imposes a critical constraint on the environmental noises, ensuring that fluctuations along the gauge orbit do not affect physical outcomes. All these results can be derived purely from symmetry considerations.

Electromagnetism in a medium has been a major area of study in physics for well over a century, resulting in an extensive body of literature. Our work is not the first to merge concepts from open quantum systems with Quantum Field Theory (QFT) — a textbook treatment can be found, for instance, in \cite{breuerTheoryOpenQuantum2002}. However, we emphasize several key differences from this standard approach. First, while \cite{breuerTheoryOpenQuantum2002} focuses on matter currents when the electromagnetic field is integrated out, we instead consider the propagation of the electromagnetic field in a medium whose dynamics is integrated out. Our aim is to characterize its self-consistent open dynamics of the two spin-one helicities of photon without detailed knowledge of the microscopical environment. Second, we avoid a gauge-fixed framework, instead emphasizing the role of gauge symmetries in open QFT. Finally, we do not restrict ourselves to thermal equilibrium, embracing a fully non-equilibrium treatment.

This article is organized as follows. In \Sec{sec:SK}, we review the main features of the Schwinger-Keldysh formalism. In \Sec{sec:scalars}, we derive some general results on the characterization of dissipative free theories. In \Sec{sec:OEM}, we apply this knowledge to the construction of Open Electromagnetism, a theory that describes the propagation of light in a medium. Our main conclusions are gathered in \Sec{sec:conclu} where the results are discussed.


\paragraph{Notations and conventions.} We work in the mostly positive signature. The $4$-momenta are given by $k^\mu = (\omega, \bmk)$, while the notation without spacetime indices $k = |\bmk|$ stands for the norm of the spatial momenta. The normal vector $n^\mu = (-1, \bm{0})$ is a timelike $4$ vector associated with a preferred timelike direction that breaks Lorentz invariance. In general, EFT coefficients are analytic functions of $ \omega$ and $ \bfk$ around the origin by locality in time and space, assuming isotropy. The convention for the Fourier transform is 
\begin{align}
    f(x-y) &=\int\frac{d^{4}k}{(2\pi)^{4}}f(\omega, \bmk)e^{ik^\alpha(x-y)_\alpha},
\end{align}
such that time derivatives generate powers of $- i \omega$ and space derivatives power of $+i \bmk$ in Fourier space
\begin{align}
\partial_\mu \to (-i\omega,+i\bfk)    \,.
\end{align}
Throughout the article, $\langle \hat{\mathcal{O}}(t) \rangle$ stands for the quantum expectation value of the quantum operator $\hat{\mathcal{O}}$ at time $t$, explicitly defined by $\Tr[\hat \rho(t)\hat{\O}(t)]$, with $\hat{\rho}(t)$ the density matrix of the system. The double bracket notation instead is a shorthand notation for the path integral average over some function of the path integral fields, which carry a label $\pm$ depending on the branch of the Schwinger-Keldysh contour on which they live,
\begin{align}
\exx{ \O[\varphi_+,\varphi_-]}\equiv \int \dd \phi \int^{\phi }_\text{I.C.} \mathcal{D}\varphi_+\int^{\phi}_\text{I.C.} \mathcal{D}\varphi_-\O[\varphi_+,\varphi_-] e^{iS_{\mathrm{eff}}[\varphi_+,\varphi_-]}\,,
\end{align}
where ``I.C." denotes some initial conditions.


\section{Schwinger-Keldysh formalism}\label{sec:SK}

The Schwinger-Keldysh formalism \cite{Schwinger:1960qe, Keldysh:1964ud} aims at computing expectation values of operators in far-from-equilibrium systems and/or open systems. In this article, we follow the EFT philosophy and aim at providing the most general description of the system while remaining as agnostic as possible about the environment. We assume separation of scales between system and environment such that interactions mediated by the environment can be modeled  by a finite number of local, possibly dissipative, interactions of the system. We will therefore refrain from a detailed description of the environment, trading it for a self-consistent description of the system interactions. This is similar in spirit to the Lindblad description of open quantum system \cite{Lindblad:1975ef, 10.1063/1.522979} where the focus is made on the derivation of a well-defined open dynamics, avoiding microphysical modeling of the environment. For concreteness, we assume that the environment is homogeneous, isotropic, and invariant under time translation. Moreover, we neglect the backreaction of system on the environment. There are there three steps in constructing local EFTs: i) choosing the degrees of freedom; ii) determining the symmetries of the theory, iii) keeping a finite number of operators in a radiatively stable power counting scheme.

In this work, we consider the latter situation where a system of interest interacts with an unspecified environment. The setup assumes a unitary $\{$system $+$ environment$\}$ evolution. Starting in a pure state and upon tracing out the environment, the system eventually ends up in a mixed state described by the density matrix $\hat \rho(t)$. In this open quantum system, equal-time expectation values of fundamental operators are schematically computed by
\begin{align}\label{eq:expect}
\ex{\hat \O(t)}=\Tr[\hat \rho(t)\hat{\O}(t)]=\int \dd\phi \, \dd \phi' \, \rho_{\phi \phi'}(t) \bra{\phi'}\hat \O(t)\ket{\phi}\,,
\end{align}
with 
\begin{align}\label{rhomatrix}
\rho_{\phi\phi'}(t)\equiv \bra{\phi}\hat \rho(t)\ket{\phi'}=\int^{\phi} \mathcal{D}\varphi_{+}\int^{\phi'}\mathcal{D} \varphi_{-}\,e^{iS_{\mathrm{eff}}[\varphi_{+},\varphi_{-}]}\,.
\end{align}
$S_{\mathrm{eff}}$ is a functional of the fields that we will refer to as \textit{open functional}. This is the sum of terms describing the unitary dynamics of the system plus an influence functional describing effects mediated by the environment \cite{FEYNMAN1963118}. Notice that $ \dd\phi$ denotes an average over boundary conditions at some time-slice $ t$, while $ \mathcal{D}\pip$ and $ \mathcal{D}\pim$ denote path integrals over histories \cite{Boyanovsky:2015xoa}. The initial conditions, which we left implicit in the formulae, are specified in terms of an initial density matrix, which we will always to be a pure state in the infinite past. These expressions are nothing but the field basis representation of the density matrix. In the rest of this work, we will be interested in operators that are diagonal in the field basis, namely $\bra{\phi}\mathcal{O}\ket{\phi'}\propto \delta(\phi-\phi')$. This is the case for the product of fields $\phi$ at different spacetime points, but it is not the case if one includes their momentum conjugate. Hence we will be interested in the \textit{diagonal elements} of the density matrix $\rho_{\phi \phi}(t)$ appearing in \Eq{eq:expect} for which $\phi' = \phi$.


\subsection{Non-equilibrium constraints}\label{subsec:NEQconst}

\begin{figure}[tbp]
    \centering
    \includegraphics[width=1\textwidth]{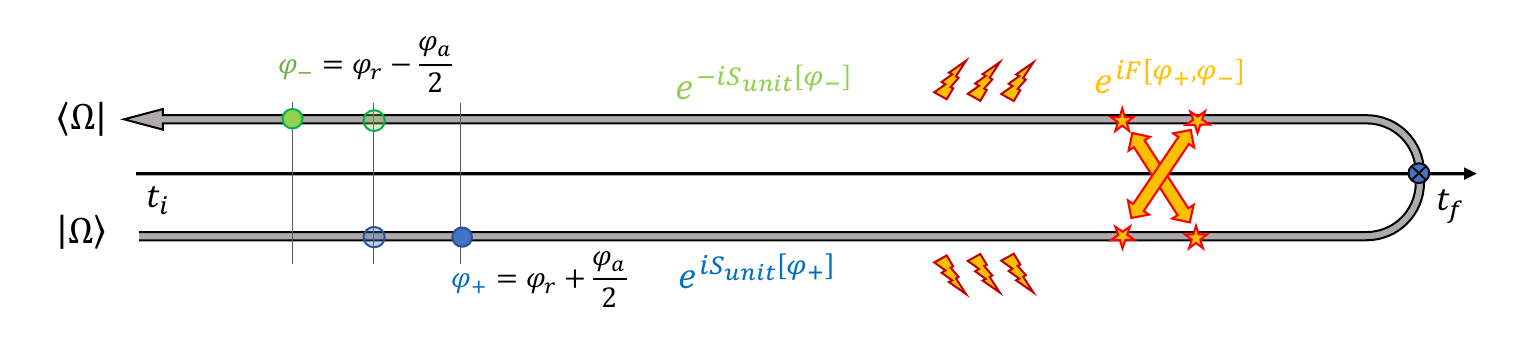}
\caption{Illustration of the Schwinger-Keldysh path integral, where time is running from left to right in both contours and the arrow represent path ordering (time ordering in $\ket{\Omega}$ and anti-time-ordering in $\bra{\Omega}$). We consider an initially pure state $\ket{\Omega}\bra{\Omega}$ at initial time $t_i$. The unitary time evolution, which preserves the purity of the state, is the same with opposite sign on each branch of the path integral, captured by $S_{\mathrm{unit}}$. Dissipative and stochastic effects are then captured by $F$ which has no unitary counterpart and cannot be captured through a single-branch contour.}
    \label{fig:contour}
\end{figure}

By definition, $\hat\rho$ obeys the following constraints \cite{breuerTheoryOpenQuantum2002}
\begin{align}\label{eq:NEQ}
\Tr \hat\rho&=1\,, & \hat{\rho}^{\dagger}&=\hat\rho\,, & \hat\rho&\geq0\,,
\end{align}
where the last is a shorthand notation for $ \bra{\psi}\hat \rho \ket{\psi}\geq 0$ for all $ \ket{\psi}\in \H$. These conditions are crucial to ensure that the quantum mechanical formalism makes meaningful statistical predictions. These three defining properties of a density matrix impose additional constraints on the functional $ S_{\mathrm{eff}}$ \cite{Calzetta:2008iqa, Liu:2018kfw} that are respectively\footnote{Note that these properties imply that the diagonal elements $ \rho_{\phi\phi}$ are real.}
\begin{align}
    S_{\mathrm{eff}} \left[\varphi_+,\varphi_+\right] &= 0\,,  \\
    S_{\mathrm{eff}} \left[\varphi_+,\varphi_-\right] &= - 	S^*_{\mathrm{eff}} \left[\varphi_-,\varphi_+\right] \,, \label{herm}\\
    \Ima S_{\mathrm{eff}} \left[\varphi_+,\varphi_-\right] &\geq 0 \,.
\end{align}
For us $S_{\mathrm{eff}}$ is a functional that integrates an integrand over the whole of space, for example $\mathbb{R}^3$, and over some interval of time. The above conditions have to be satisfied for any time interval and hence imply the vanishing of the integrand at every spacetime point $(t,\bmx)$ where $\varphi_a(t,\bmx)=0$. Sometimes it might be convenient to break up the functional $ S_{\mathrm{eff}}$ into a part that describes unitary evolution on a normalizable pure state and the rest, that is\footnote{One may further break $F[\varphi_{+},\varphi_{-}] $ into 
\begin{align}
    F[\varphi_{+},\varphi_{-}] = F_{\mathrm{mix}}[\varphi_+, \varphi_-] +  F_{\mathrm{no-mix}}[\varphi_+] +  F_{\mathrm{no-mix}}[\varphi_-],
\end{align}
with terms that mix the two branches of the path integral such that $F_{\mathrm{mix}}[\varphi_+, 0] = F_{\mathrm{mix}}[0, \varphi_-] = 0$ and a non-mixing piece (yet non-unitary) 
\begin{align}
    F_{\mathrm{no-mix}}[\varphi]\equiv \frac{1}{2} \left(  S_{\mathrm{eff}}[\varphi,0]+S_{\mathrm{eff}}[0,\varphi]\right),
\end{align} 
obeying $F_{\mathrm{no-mix}}[\varphi] =-F_{\mathrm{no-mix}}[\varphi]^{*}$. As it will appear clearly in the Keldysh basis, the physical interpretation of $F[\varphi_{+},\varphi_{-}]$ often relies on the combined effect of $F_{\mathrm{mix}}[\varphi_+, \varphi_-]$ and $F_{\mathrm{no-mix}}[\varphi]$, which explains why we have not further separated these terms.}
\begin{align}\label{eq:splitting}
S_{\mathrm{eff}}[\varphi_{+},\varphi_{-}]&= S_{\mathrm{unit}}[\varphi_{+}]-S_{\mathrm{unit}}[\varphi_{-}] + F[\varphi_{+},\varphi_{-}]\,.
\end{align}
Indeed, if one sets $ S_{\mathrm{unit}}[0]=\dot{S}_{\mathrm{unit}}[0]=0$ by subtracting a constant and similarly for $F$, one can find the defining relations
\begin{align}
S_{\mathrm{unit}}[\varphi]&\equiv \frac{1}{2}\left(  S_{\mathrm{eff}}[\varphi,0]-S_{\mathrm{eff}}[0,\varphi]\right).
\end{align}
The condition \eqref{herm} of $ \hat\rho^{\dagger}= \hat\rho$ then implies
\begin{align}
S_{\mathrm{unit}}[\varphi]&=S_{\mathrm{unit}}^{*}[\varphi] \,.
\end{align}
The decomposition is summarized in \Fig{fig:contour}. $F$ is often called the Feynman-Vernon influence functional and encodes the effects of the environment on the system \cite{FEYNMAN1963118}. We will not derive $F$ from an explicit model of the environment. Instead, our goal here is to model $F$ in the most generic way possible by assuming a set of symmetries and, most importantly, a separation of scales that ensures locality in time and space in the system sector. \\

We often work in the Keldysh basis 
\begin{align}
    \varphi_r = \frac{\varphi_+ + \varphi_-}{2},\qquad \varphi_a = \varphi_+ - \varphi_- \qquad \Leftrightarrow \qquad \varphi_+ = \varphi_r + \frac{\varphi_a}{2}, \qquad \varphi_- = \varphi_r - \frac{\varphi_a}{2} .\label{eq:RAbasis}
\end{align}
The retarded and advanced components $\varphi_r$ and $\varphi_a$, which respectively correspond to the mean and the difference of the field inserted along each branch of the path integral, turn out to conveniently organise the perturbative expansion \cite{Kamenev:2009jj}. In this basis, the above non-equilibrium constraints become 
\begin{align}
S_{\mathrm{eff}} \left[\varphi_r,\varphi_a = 0\right] &= 0 \,,\label{eq:norm} \\
S_{\mathrm{eff}} \left[\varphi_r,\varphi_a\right] &= - 	S^*_{\mathrm{eff}} \left[\varphi_r,-\varphi_a\right] \,,\label{eq:herm} \\
\Ima S_{\mathrm{eff}} \left[\varphi_r,\phi_a\right] &\geq 0. \label{eq:pos}
\end{align} 
While this basis is convenient to make manifest the causality structure of the theory \cite{kamenev_2011} and to understand the structure of non-unitary operators \cite{Salcedo:2024smn}, it renders much harder the identification of a unitary subset, for which the $+/-$ basis remains the best option.


\subsection{Boundary conditions: a toy model} \label{subsec:BCs}

To explore the basic features of this formalism, let us first consider a single scalar degree of freedom $\varphi$ evolving in flat space according to 
\begin{align}\label{eq:BM}
    S_{\mathrm{eff}}[\phir,\phia] = \int \dd^4x \left(\dphir \dphia - c_s^2 \partial_i \phir \partial^i \phia + \gamma \dphir \phia + i \beta \phia^2\right)
\end{align}
where a dot represents a time derivative. We stress that we are not assuming that the system or the medium are in thermodynamical equilibrium. In particular $\beta$ is not in general related to the inverse temperature. In other words, our Schwinger-Keldysh path integral prepares a density matrix that is not thermal. If desired, one could impose additional constraints, known as Kubo–Martin–Schwinger (KMS) conditions (see \cite{Kubo:1966fyg, PhysRev.115.1342} and \cite{Hongo:2018ant} for a review), to specify our construction to the thermal case. We do not follow this approach in this work because our eventual goal is to develop this formalism for problems in cosmology and gravity where thermalization does not necessary take place.

First, note that the first two terms can be written in a factorized form in the original $+/-$ basis,
\begin{align}
    S_{\mathrm{unit}}[\varphi_+] - S_{\mathrm{unit}}[\varphi_-] \qquad \mathrm{with} \qquad S_{\mathrm{unit}}[\varphi] = \frac{1}{2} \int \dd^4x \left[\dot{\varphi}^2 - c_s^2 (\partial_i \varphi)^2 \right],
\end{align}
and hence represent the unitary evolution. 
Conversely, the last two terms have no analogue in the unitary case and encode the dissipative and diffusive effects of $F[\varphi_+,\varphi_-]$ characteristic of the open dynamics. Note that both $\gamma$ and $\beta$ should be real, in accordance with the conditions above. 

Let us first discuss the first non-unitary contribution $\gamma \dphir \phia$ appearing in \Eq{eq:BM}. This dissipative operator is crucial in describing the loss of energy of the system $\varphi$ into its surrounding environment. In the original $+/-$ basis, it contains the boundary term
\begin{align}
    S_{\mathrm{unit}}[\varphi_+] - S_{\mathrm{unit}}[\varphi_-] \qquad \mathrm{with} \qquad S_{\mathrm{unit}}[\varphi] = -\frac{1}{4} \int \dd^4 x \left(\frac{\dd}{\dd t} \left[\varphi^2\right] \right),
\end{align}
together with a mixing between the two branches of the path integral 
\begin{align}
    F[\varphi_+,\varphi_-] = - \frac{1}{2} \int \dd^4 x \left(\dot{\varphi}_+ \varphi_- - \varphi_+ \dot{\varphi}_- \right).
\end{align}
Upon specifying the boundary conditions of the diagonal density matrix element appearing in \Eq{eq:expect}
\begin{align}
    \varphi_+(\bmx, t_0) = \varphi_-(\bmx, t_0) = \phi(\bmx),
\end{align}
it is clear that the boundary term does not contribute to the diagonal elements of $\hat \rho$. On the contrary, the mixing term $F[\varphi_+,\varphi_-] $ has no reason to vanish. This effect has no unitary counterpart and is precisely responsible for damping of $\phi$ fluctuations due to dissipation. To make this fact manifest, we now need to derive the equations of motion.  

Before doing this, let us comment on the fact that for a single scalar field, there exist four related options to introduce dissipation, all differing by boundary conditions. These options are 
\begin{align}
(1a):~ \quad \dot\varphi_{r}\varphi_{a} \,, \qquad & (1b): \quad \varphi_{r}\dot\varphi_{a} \,, \\
(2a): \quad \dot\varphi_{+}\varphi_{-} \,, \qquad & (2b): \quad \varphi_{+}\dot\varphi_{-} \,.
\end{align}
In Minkowski, (1a) is related to (1b) by integration-by-part and the same applies to (2a) and (2b). Also, in Minkowski, (1a) is the same as (2a) up to boundary terms. Consequently, all four options are the same up to boundary terms. Conversely, in de Sitter (or any FLRW spacetime), any two of these terms differ by boundary terms as well as mass terms (due to the integration by part acting on the scale factors).\footnote{It is interesting to ask how the theory for a Goldstone $ \pi $ selects this mass term. In the $ \pi_{r,a} $ variables, we know a mass term is forbidden, since $  \pi_{r}\pi_{a} $ is not shift invariant in $  \pi_{r} $. Indeed, the flat space dispersion relation is special in that it displays a gapless mode for $  k\to 0 $ (as well as a gapped mode) obtained from \cite{Salcedo:2024smn}.
In $  \pi_{\pm} $ coordinates, the same result is found by noticing that $  \dot \pi_{\pm} \pi_{\mp} $ is actually allowed even if it is not manifestly shift symmetric. In de Sitter, the Goldstone $ \pi  $ does acquire a mass, but we neglected it in \cite{Salcedo:2024smn} because it is of the same order as the mixing with gravity (hence slow-roll suppressed), which we neglected.}

 
\subsection{Saddle-point approximation}\label{ssec:saddle}

The structure of the Schwinger-Keldysh path integral in the Keldysh basis is really peculiar and one cannot stress enough how important the boundary conditions are. The boundary conditions for $  \varphi_{a} $ are those responsible for $  \varphi_{a} $ not propagating, despite the fact that $  \varphi_{a} $ appears in the quadratic action in a way that is very similar to $  \varphi_{r} $. If it was not for the boundary conditions, the theory would inevitably suffer from an instability.\footnote{To see this, imagine to incorrectly assume that $  \varphi_{a} $ has standard initial conditions (as opposed to satisfying a boundary condition problem) and integrate it out. Its equation of motion is given by $  \varphi_{a}=\Box \varphi_{r} $, which, once substituted back into the action would give terms as $  \varphi_{r}\Box^{2}\varphi_{r} $. This theory is the iconic example of a ghost, as seen from the fact that the propagator can be partial fractioned into two terms with opposite signs (this is best seen introducing a small mass, which can be sent to zero at the end). Another way to see this is that $  \varphi_{+} $ and $  \varphi_{-} $ have opposite sign kinetic terms (since we have $iS_{\mathrm{unit}}[\varphi_{+}]-iS_{\mathrm{unit}}[\varphi_{-}]  $) and one of the two would end being a ghost.} Because of the boundary conditions, $  \varphi_{a} $ does not propagate and the theory is healthy. However, something peculiar happens as a byproduct: the $  \varphi_{a} $ path integral does not admit a stationary phase! This is because, in the presence of fluctuations $  i \beta \varphi_{a}^{2} $, the stationary phase would be $  \varphi_{a}\sim \Box \varphi_{r} $. This relation is manifestly incompatible with the boundary conditions that $  \varphi_{r} $ is anything but $  \varphi_{a} $ vanishes at $  t=t_{0} $. In other words, \textit{if $  \varphi_{a} $ appears non-linearly its path integral does not admit a stationary phase}.

Things would change if $  \varphi_{a} $ appeared linearly. Then, its stationary phase would be some equation of motion for the \textit{other} fields, not involving $  \varphi_{a} $, and that would always be compatible with the vanishing of $  \varphi_{a} $ on the boundary. This is indeed the magic that ensues from the Hubbard-Stratonovich (HS) transformation \cite{Stratonovich1957,Hubbard1959}. Upon introducing an auxiliary noise field $ \xi   $, the open functional becomes \textit{linear} in $  \varphi_{a} $. It is only at this point that the path integral in $  \varphi_{a} $ is well approximated by a stationary phase and it is only now that $  \delta S/ \delta \varphi_{a} $ can be interpreted as an equation of motion, namely the Langevin equation. Explicitly, the key trick is the mathematical identity
\begin{equation}\label{eq:HStrick}
    \exp\left(-\int d^{4}x \beta\phia^{2}\right)=\mathcal{N}_{0}\int\mathcal{D}\xi\;\exp\left[\int d^{4}x \left(-\frac{\xi^{2}}{4\beta}+i\xi\phia\right)\right],
\end{equation}
with $\mathcal{N}_{0}$ being the normalisation constant. After this transformation, the path integral in $  \varphi_{a} $ is well approximated by a stationary phase which generates the Langevin equation
\begin{align}\label{eq:langevin}
    \ddot{\varphi}_r + \gamma \dphir + c_s^2 k^2 \phir = \xi.
\end{align}
The new variable $\xi$, satisfying $\ex{\xi(x)}=0$, behaves as a Gaussian field with a prescribed two-point function
\begin{equation}\label{eq:noisecontact}
    \langle\xi(x) \xi(y)\rangle=2\beta \delta(x-y).
\end{equation}


\subsection{The role of $\varphi_a^{2n+1}$ terms in a unitary theory} 

The discussion above is nice as it gives a way to connect to the classical dynamics via the Langevin equation. However there is still one puzzle on which we wish to comment. When writing a general unitary interacting theory in the $\varphi_{a,r}$ basis we find terms such as $  \varphi_{a}^{2n+1} $. What is the role of these terms? It is not easy to see how they contribute to the Langevin equation because we have to perform more complex version of the Hubbard-Stratonovich transformation, as done in \cite{Salcedo:2024smn} for $\varphi_a^3$ (see also \cite{Chakrabarty:2019aeu}). Moreover, we know that the theory is unitary, so it should have a conservative (no dissipation) and deterministic (no noise) equation of motion. To better understand the role of these terms, consider the simple toy model of a massive scalar with $  \lambda \varphi^{3} $ interaction
\begin{align}
S_{\mathrm{eff}}[\phir,\phia] = S_{\mathrm{unit}}[\varphi_+] - S_{\mathrm{unit}}[\varphi_-] \quad \mathrm{with} \quad S_{\mathrm{unit}}[\varphi] = \int \dd^4x \left[ -\frac{1}{2}\left(\partial_{\mu}\varphi\right)^{2}-\frac{1}{2}m^{2}\varphi^{2} -\lambda \varphi^{3}\right], \label{eq:unitarySphi}
\end{align}
that is 
\begin{align}
S_{\mathrm{eff}}[\phir,\phia] &=\int \dd^4x \left(-\partial_{\mu}\varphi_{a}\partial^{\mu}\varphi_{r}-m^{2}\varphi_{a}\varphi_{r}-3\lambda \varphi_{a}\varphi_{r}^{2}-\frac{\lambda}{4} \varphi_{a}^{3} \right)\,.\label{eq:unitaryIphi}
\end{align}
If we vary $S_{\mathrm{eff}}$ with respect to $\varphi_a$, the terms linear in $  \varphi_{a} $ precisely generate the classical equations of motion, with all the right coefficients. At this point, it is not very clear what the goal in life is of the additional term $  \varphi_{a}^{3} $. In \cite{Salcedo:2024smn}, Appendix D, we computed the flat-space contact bispectrum and checked that it is essential to account for both the $  \varphi_{a} \varphi_{r}^{2} $ and the $  \varphi_{a}^{3} $ contributions to get the right vacuum unitary result. In other words, neglecting $\varphi_a^3$ would give the \textit{wrong} tree-level three point function. In general, when considering a ``vacuum" initial state, one should not discard cubic and higher operators in the advanced field, contrarily to the often justified approach in the case of highly occupied states \cite{Radovskaya:2020lns, 2016RPPh...79i6001S}. 

We make a quick detour to highlight how to recover the standard vacuum results of the cubic unitary theory. The observables we aim to match between \eqref{eq:unitarySphi} and \eqref{eq:unitaryIphi} are the in-in equal time correlators of the field $\hat{\phi}(x)$. In the case of \eqref{eq:unitarySphi}, these quantities are computed via a field theoretical wavefunctional, then using the Born rule. In the case of \eqref{eq:unitaryIphi}, we would use either the Feynman rules for the Schwinger-Keldysh formalism or the Langevin equation. In both cases, it is required to introduce an $\epsilon$ prescription in \eqref{eq:unitaryIphi} to derive the Keldysh propagator when performing the Schwinger-Keldysh path integral.
This prescription is \cite{2016RPPh...79i6001S}
\begin{equation}
    S_{\mathrm{eff}}[\phir,\phia] =\int \dd^4x \left(-\partial_{\mu}\varphi_{a}\partial^{\mu}\varphi_{r}-2\epsilon\varphi_{a}\dot{\varphi}_{r}-(m^{2}+\epsilon^{2})\varphi_{a}\varphi_{r}+i\epsilon \fini\varphi_{a}^{2}-3\lambda \varphi_{a}\varphi_{r}^{2}-\frac{\lambda}{4}\varphi_{a}^{3} \right)\,.\label{eq:epsIphi}
\end{equation}
The first term featuring $\epsilon$ corresponds to an arbitrarily small dissipation that enforces the convergence of the time integrals as $t\to-\infty$ for the retarded propagator of the Langevin equation. The $\epsilon$ correction to the mass term is present to ensure that the left-hand side Langevin equation is deformed as
\begin{equation}
    \left[(\partial_{0}+\epsilon)^{2}+m^{2}\right]\varphi=\xi+...
\end{equation}
which yields the small imaginary displacement of the real-axis pole in the frequency plane for the computation of the retarded and advanced propagators. At last, the $i\epsilon \fini\varphi_{a}^{2}$ term corresponds to the particle population of the state at initial time on which we are computing the in-in correlators. It directly relates to the initial occupation of the state through $\fini = 2 n_0 + 1$ with 
\begin{align}
    n_0(E_k) = \frac{1}{e^{E_k/T}-1},
\end{align}
for a bosonic system initially prepared in a thermal state at temperature $T$, with $E_k = \sqrt{k^2 + m^2}$ in this case. The vacuum limit is recover when taking $T\rightarrow 0$. The Hubbard-Stratonovich trick in this case works just as above, that is
\begin{equation}
    \text{exp}\left[-\int d^{4}x\;\epsilon \fini\varphi_{a}^{2}\right]=\int[\mathcal{D}\xi]\,\text{exp}\left[\int d^{4}x\left(-\frac{\xi^{2}(x)}{4\epsilon \fini}+i\varphi_{a}\xi\right)\right].
\end{equation}

The introduction of these terms allows us to either write a Langevin equation from which to compute in-in equal time correlators, or equivalently to derive the vacuum Keldysh propagator. These will at first sight depend on $\epsilon$ and, in the limit $\epsilon\to0^{+}$, yield finite results which correspond to the standard correlators computed through the unitary action \eqref{eq:unitarySphi}. Explicitly, the full non-linear Langevin equation is
\begin{equation}\label{eq:NLLang}
\left[(\partial_{0}+\epsilon)^{2}-\nabla^{2}+m^{2}\right]\varphi(x)=\xi(x)+ \frac{\lambda}{4\epsilon^{2}f^{2}}\xi^{2}(x)-3\lambda\varphi^{2}(x),
\end{equation}
where the $\xi^{2}(x)$ term comes from the advanced cubic operator $\varphi_a^3$, see Section 5 of \cite{Salcedo:2024smn} for details.\footnote{In the presence of $\varphi_{a}^{3}$ operator, the Hubbard–Stratonovich trick is modified to \cite{Salcedo:2024smn} 
\begin{align}
&\exp\left[-\int d^{4}x\left(\epsilon f +i \frac{\lambda}{4} \varphi_a\right)\varphi_a^{2}\right]=\mathcal{N}(\varphi_a)\int[\mathcal{D}\xi] \exp\left[\int d^{4}x \left(-\frac{\xi^2}{4\epsilon f +i \lambda \varphi_a}+i\varphi_a \xi\right)\right].
\end{align}
Expanding at leading order in $\varphi_a$ both the denominator and the normalisation constant, we obtain a term of the form $\xi^2\varphi_a$, that is
\begin{align}
&\exp\left[-\int d^{4}x\left(\epsilon f +i \frac{\lambda}{4} \varphi_a\right)\varphi_a^{2}\right] \approx\mathcal{N}\int[\mathcal{D}\xi] \exp\left[\int d^{4}x \left(-\frac{\xi^{2}}{4\epsilon f} + i\xi\varphi_a + i\frac{\lambda}{4\epsilon^2 f^2} \xi^2 \varphi\right) \right].  
\end{align} 
The $\xi^{2}\varphi_a$ term enters the Langevin equation through the $\xi^2$ term in \Eq{eq:NLLang}.
}

\vspace{0.5in}
We first recover the propagator for Schwinger-Keldysh perturbation theory, for which one needs to take the limit of $\epsilon\to0$ on the convolution of two retarded propagators:
\begin{equation}
    G^{K}(x,y)=\lim_{\epsilon\to0^{+}}\left[2\epsilon\int d^{4}z f(z)G^{R}(x,z)G^{R}(y,z)\right],
\end{equation}
where $G^{R}(x,z)$ is the retarded propagator of the Green's equation:
\begin{equation}
    \left[(\partial_{0}+\epsilon)^{2}-\nabla_{x}^{2}+m^{2}\right]G^{R}(x,z)=\delta(x-z).
\end{equation}
The finiteness of the Keldysh propagator arises from a cancellation between the $\epsilon$ power from the noise term $i\epsilon \fini\varphi_{a}^{2}$ and the $\epsilon^{-1}$ divergence in the $z$ integral that is regularized thanks to the dissipation term $2\epsilon\varphi_{a}\dot{\varphi}_{r}$.

To compute in-in correlators, ordering of limits is crucial. At finite $\epsilon$, we first perform the calculations in perturbation theory for $\lambda$. Then, once we have computed the in-in correlator, we can take the $\epsilon\to0$ limit. While the amplitude of the noise power spectrum vanishes for $\epsilon\to0^{+}$, that is $\langle\xi\xi\rangle\propto \epsilon \fini$, one can take a field redefinition for $\xi$ to $\xi^{\text{NG}}$ 
\begin{equation}
    \xi^{\text{NG}}(x)=\xi(x)+ \frac{\lambda}{4\epsilon^{2}f^{2}}\xi^{2}(x),
\end{equation}
to find that the three-point function of $\xi^{\text{NG}}$ is non-zero in the $\epsilon\to0^{+}$ limit
\begin{equation}
    \langle \xi^{\text{NG}} \xi^{\text{NG}} \xi^{\text{NG}}\rangle\propto  \frac{\lambda}{4\epsilon^{2}f^{2}}\langle \xi \xi \xi^{2}\rangle\propto \lambda.
\end{equation}
This brings light on the finite contributions of the $\varphi_a^3$ terms, interpreted as a non-Gaussian noise term, to the vacuum in-in correlators. 

If the $\epsilon$ prescription is so crucial to impose initial conditions and recover vacuum expectation values, the reader may wonder why it is so rarely discussed in the non-equilibrium literature. The reason is that whenever the open system experiences a small but finite dissipation $\gamma \dot{\varphi}_r \varphi_a$, initial conditions get erased after a finite time. Asymptotic observables do not depend on the initial occupation of the state and information about it has been lost throughout the dynamics. At a mathematical level, one can see that $\gamma$ induces a pole in the frequency plane that guarantees the convergence of the integral contour used to compute the retarded and advanced Green's function, such that the small $\epsilon$ deformation becomes unnecessary in practice.


\subsection{Field redefinitions} 

One of the main features of the Schwinger-Keldysh formalism is the introduction of an advanced component $\varphi_{a}$ that is non-dynamical, its boundary conditions being entirely fixed. Moreover, if we exclusively look at the diagonal components of the density matrix we know that $\varphi_a=0$ at the ``boundary" , i.e. where the Schwinger-Keldysh path integral turns around. Because of this, we are allowed to perform field redefinitions that mix the retarded and the advanced components as long as those respect the boundary conditions. Allowed field redefinitions are, for instance,
\begin{align}
    \varphi_{r}&\rightarrow\varphi_{r}+F_{r}(\varphi_{r},\varphi_{a}),\label{eq:fieldref}\\
    \varphi_{a}&\rightarrow\varphi_{a}+F_{a}(\varphi_{r},\varphi_{a}),\nonumber
\end{align}
as long as $F_{r,a}(\varphi_{r},\varphi_{a}=0)=0$. Note that we have the freedom to change the boundary condition of the retarded field as eventually we integrate over all possible boundary conditions when computing equal time in-in correlators of $\phi$, see \Eq{eq:expect}. One of the main uses of field redefinitions is the removal of terms from the action that are proportional to the equations of motion. For a general open theory, it is possible for these terms to appear in the open effective functional in a non-unitary fashion, that is by mixing the two branches of the path integral. As an example, we could consider the case of
\begin{equation}
    S_{\text{eff}}[\varphi_{r},\varphi_{a}]=\int d^{4}x\left[\varphi_{a}\hat{\mathcal{O}}_{\text{EOM}}\varphi_{r}+...+\frac{g}{\Lambda}\varphi_{a}\hat{\mathcal{O}}_{\text{EOM}}\varphi_{a}^{2}\right],
\end{equation}
where $g$ is some dimensionless coupling and $\Lambda$ is some energy scale. The term $\varphi_{a}\hat{\mathcal{O}}_{\text{EOM}}\varphi_{a}^{2}$ may not be accompanied by a corresponding $\mathcal{O}(\varphi^{2}_r\varphi_{a})$ term that would make it unitary. 
However, the field redefinition
\begin{equation}
    \varphi_{r}\rightarrow\varphi_{r}-\frac{g}{\Lambda}\varphi_{a}^{2}
\end{equation}
is enough to remove the non-unitary term proportional to the equations of motion. This term in particular has the advantage of not changing the boundary condition of $\varphi_{r}$. 


\subsection{Dispersion relations}\label{subsec:dispreleg}

We have illustrated the simplest example \eqref{eq:BM} encompassing local dissipation and noise. The rest of the paper will feature a gradual increase in complexity. First we will considering more than one dynamical degree of freedom and then we include constrained fields and finally gauge symmetries. Our discussion culminates with the study of electromagnetism in a medium, which we dub \textit{open electromagnetism}.
Many of these situations will require a neat accounting of the number of propagating degrees of freedom and a discussion of stability conditions. For this reason, it is instructive to go back to the simplest situation of \Eq{eq:BM}. In Fourier space, after the Hubbard-Stratonovich trick, the effective functional is rewritten as
\begin{align}\label{eq:disppre}
    S_{\mathrm{eff}} = \int \frac{\dd \omega}{(2\pi)} \int \frac{\dd \bmk}{(2\pi)^3} \left[\varphi_r \left(- \omega^2 + c_s^2 k^2- i \gamma \omega \right)\varphi_a + i \beta \varphi_a^2\right].
\end{align}
The dispersion relation 
\begin{align}\label{eq:disprelref}
    \omega^2 + i \gamma \omega - c_s^2 k^2 = 0
\end{align}
characterises the propagation in the system. Explicitly, it is the fact that \Eq{eq:disprelref} admits non-trivial solutions for the frequencies $\omega(k,\gamma)$ that ensures the existence of a dissipative degree of freedom. In this case, the solutions are 
\begin{align}\label{eq:soldispeg}
   \omega_\pm = -i\frac{\gamma}{2}\pm \sqrt{c_s^2 k^{2}-(\gamma/2)^{2}},
\end{align}
which have both a real and imaginary part whenever $2c_s^2 k^2 > \gamma$. Note that in the limit $k \rightarrow 0$ with $\gamma$ finite, the dispersion relation is purely imaginary. We notice a gapless mode, corresponding to $\omega_+ \rightarrow 0$ for $k \rightarrow 0$ and a gapped mode associated to $\omega_- \rightarrow i \gamma$ for $k \rightarrow 0$. Importantly, for $\gamma > 0$ both modes have a negative imaginary part
\begin{align}
\gamma > 0 \qquad  \Rightarrow \qquad \Ima \omega_\pm < 0,
\end{align}
so that the retarded Green's function 
\begin{align}
    G^{R} (k; \omega) &= - \frac{1}{\omega^2 + i \gamma \omega - c_s^2k^2}= -\frac{1}{(\omega_- - \omega_+)} \left[ \frac{1}{\omega- \omega_-} - \frac{1}{\omega- \omega_+}   \right],
\end{align}
is manifestly convergent, that is
\begin{align}\label{eq:retardedM4dissip}
G^{R} (k; t - t') &= \int \frac{\dd \omega}{2\pi} \ee^{i \omega (t-t')} G^{R} (k; \omega) = - \frac{\sin \left[E_k^{\gamma} (t-t')\right]}{E_k^{\gamma}} \ee^{- \frac{\gamma}{2}(t-t')}\theta\left(t-t'\right),
\end{align}
where we defined $E_k^{\gamma} = \sqrt{c_s^2 k^{2}-(\gamma/2)^{2}}$. Note that the causality property encoded in the above Heaviside theta function indeed implies analyticity of $G^R$ in the upper-half complex frequency plane.
Therefore, $\gamma > 0$ ensures causality and stability, via late-time convergence. 

Note that the noise term $ i \beta \varphi_a^2$ appearing in \Eq{eq:disppre} does not affect the propagation of the dissipative degree of freedom \cite{kamenev_2011}. Indeed, the noise sources the dynamics of the system but does not directly transform the propagation of information within it (said it differently, the former relates to the Keldysh function whereas the latter to the Green's functions). For this reason, in the following, we focus in terms linear in the advanced component to discuss the number of propagating degrees of freedom.


\subsection{Path ordering}\label{subsec:path}

Ultimately, the path integral aims at computing expectation values of observables. To achieve this task, we need to acknowledge what the path integral is actually computing. In this Section, we discuss the link between path integral and quantum operators. We do so by establishing the relation between path ordering both in the in-out and the in-in contour. 
\subsubsection*{In-out contour}
Our starting point is the transition from a state $ \ket{\phi_{i}}$ at time $ t_i$ to a state $ \ket{\phi_{f}}$ at $ t_f$. In the Schr\"odinger picture this is 
\begin{align}
\bra{\phi_{f}}e^{-i\int^{t_f}_{t_i}H dt}\ket{\phi_{i}}\equiv \bra{\phi_{f}}U(t_f,t_i)\ket{\phi_{i}} \,.
\end{align}
Following \cite{Polchinski:1998rq}, we denote this by first going to the Heisenberg picture, where $ 
\phi(t)=U^{\dagger}\phi(t_i)U$. Then, we define the field operator eigenstates
\begin{align}
\hat \phi(t)\ket{\phi,t}=\phi(t)\ket{\phi,t}\,.
\end{align}
Even if we work in the Heisenberg picture, where states don't evolve, these states depend on time because they are defined as eigenstates of a different operator $ \hat \phi(t)$ at each time $ t$. Their time dependence can be made explicit by writing
\begin{align}
\ket{\phi,t}=U(t,t_i)\ket{\phi,t_i}\,.
\end{align}
The key relation is that the transition amplitude can be computed by a path integral
\begin{align}
\braket{\phi_{f},t_f|\phi_{i},t_i}=\int_{\phi_{i},t_i}^{\phi_{f},t_f}[\mathcal{D}\varphi]e^{iS}\,,
\end{align}
where $ S$ is some functional that depends on the theory, and the integral is over all possible histories or ``paths'' from $ t_i$ to $ t_f$ that start and end with the configurations $ \phi_{i}$ and $ \phi_{f}$, respectively. The bridge between the path integral and the operator formalism is the following cutting/sewing relation
\begin{align}
\int_{\phi_{i},t_0}^{\phi_{f},t_f}[\mathcal{D}\varphi]e^{iS}=\int d\phi \left[ \int_{\phi_{i},t_i}^{\phi,t}[\mathcal{D}\varphi]e^{iS} \right]\left[ \int_{\phi,t}^{\phi_{f},t_f}[\mathcal{D}\varphi]e^{iS} \right]\,,
\end{align}
where the intermediate integral over $ d\phi$ represents a sum over the whole Hilbert space. Hence this relation is nothing but the familiar resolution of the identity,
\begin{align}
\braket{\phi_{f},t_f|\phi_{i},t_i}=\int d\phi \braket{\phi_{f},t_f|\phi,t}\braket{\phi,t|\phi_{i},t_i}\,.
\end{align}
Using this relation, we see that averaging in the path integral over $ \varphi(t)$ is the same as computing the expectation value of the corresponding operator $ \hat\phi(t)$ in the operator formalism
\begin{align}
\int_{\phi_{i},t_i}^{\phi_{f},t_f}[\mathcal{D}\varphi] \varphi(t) e^{iS}&=\int d\phi \left[ \int_{\phi_{i},t_i}^{\phi,t}[\mathcal{D}\varphi]e^{iS} \right] \phi(t)\left[ \int_{\phi,t}^{\phi_{f},t_f}[\mathcal{D}\varphi]e^{iS} \right]\\
&= \int d\phi \braket{\phi_{f},t_f|\phi,t} \phi(t)\braket{\phi,t|\phi_{i},t_i} \\
&=\int d\phi \bra{\phi_{f},t_f} \hat \phi(t)\ket{\phi,t}\bra{\phi,t}\ket{\phi_{i},t_i}= \bra{\phi_{f},t_f} \hat \phi(t)\ket{\phi_{i},t_i}\,.
\end{align}
More surprising than this is the fact that when averaging over two fields at different times, say $ \varphi(t) \varphi(t')$, one does not recover the two point function but rather the \textit{time-ordered} two point function:
\begin{align}
\int_{\phi_{i},t_i}^{\phi_{f},t_f}[\mathcal{D} \varphi] \varphi(t) \varphi(t') e^{iS}&=\bra{\phi_{f},t_f} \mathcal{T} \hat \phi(t) \hat \phi(t')\ket{\phi_{i},t_i}\,.
\end{align}
Notice that it does not matter if we write $ \varphi(t) \varphi(t')$ or $ \varphi(t') \varphi(t)$ because these are simply functions, not operators, and they commute. The fact that the result is the time ordered correlation function makes sense because it is only inside the time ordering that the operators $ \hat \phi(t)$ and $ \hat \phi(t')$ commute. This observation can be accompanied by the slogan: ``What the path integral does for a living is to computes time-ordered correlation functions''.  

\subsubsection*{In-in contour}
So far we have been thinking of a path that is a monotonic function of time and so time-ordering and path ordering are identified. However, more generally we can consider bends in the path such that sometime moving forward in the path actually moves us backwards in time. When bends, a.k.a. time folds are present, one often keeps track of what time fold the path integral fields are on by adding a label. For example in the Schwinger-Keldysh formalism, which has only two folds one doubles the fields, $ \varphi \to (\varphi_{+},\varphi_{-})$. The path integral average of fields restricted to live on a time-fold that moves backwards in time gives an anti-time-ordered correlator, while when fields live on different time-folds one finds Wightman functions, without any time ordering.
To make the following discussion more streamlined, it is useful to introduce the following shorthand notation for the path integral average over some function 
\begin{align}
\exx{ \O[\varphi_+,\varphi_-]}\equiv \int \dd \phi \int^{\phi }_\text{I.C.} \mathcal{D}\varphi_+\int^{\phi}_\text{I.C.} \mathcal{D}\varphi_-\O[\varphi_+,\varphi_-] e^{iS_{\mathrm{eff}}[\varphi_+,\varphi_-]}\,.
\end{align}
Let us now focus on the closed-time contour of the Schwinger-Keldysh formalism, which has only two timefolds, one going forwards and one backwards in time. Using the above notation we can write
\begin{align}
\exx{\varphi_{+}(t)\varphi_{+}(t')}&=\ex{\T \hat \phi(t)\hat \phi(t')}\,,\\
\exx{\varphi_{-}(t)\varphi_{-}(t')}&=\ex{\bar\T \hat \phi(t)\hat \phi(t')}\,,\\
\exx{\varphi_{+}(t)\varphi_{-}(t')}&=\ex{ \hat \phi(t')\hat \phi(t)}\,,\\
\exx{\varphi_{-}(t)\varphi_{+}(t')}&=\ex{\hat \phi(t)\hat \phi(t')}\,.
\end{align}
Using these results repeatedly it is easy to check that
\begin{align}
\exx{\varphi_{a}(t)\varphi_{r}(t')}&=\exx{\left[ \varphi_{+}(t)-\varphi_{-}(t) \right]\frac{1}{2} \left[ \varphi_{+}(t')+\varphi_{-}(t') \right]}\\
&=\frac{1}{2}\ex{\T \hat \phi(t)\hat \phi(t')}-\ex{\bar\T \hat \phi(t)\hat \phi(t')}-\ex{  \hat \phi(t)\hat \phi(t')}+\ex{  \hat \phi(t')\hat \phi(t)}\\
&=\theta(t'-t) \ex{[\hat \phi(t'),\hat \phi(t)]}\,,
\end{align}
which is the definition of the advanced propagator. Similarly one finds the reversed time argument to correspond to the retarded propagator: 
\begin{align}
\exx{\varphi_{r}(t)\varphi_{a}(t')}&=\theta(t-t') \ex{[\hat \phi(t),\hat \phi(t')]}.
\end{align}
The two-point function of the retarded propagator corresponds to the anticommutator of the field operator:
\begin{align}
    \exx{\varphi_{r}(t)\varphi_{r}(t')}&=\frac{1}{4}\exx{\left[\varphi_{+}(t)+\varphi_{-}(t)\right]\left[\varphi_{-}(t)+\varphi_{+}(t)\right]} \\
    &=\frac{1}{4}\left[\langle\mathcal{T}\hat{\phi}(t)\hat{\phi}(t')\rangle+\langle\hat{\phi}(t)\hat{\phi}(t')\rangle+\langle\hat{\phi}(t')\hat{\phi}(t)\rangle+\langle\bar{\mathcal{T}}\hat{\phi}(t)\hat{\phi}(t')\rangle\right]\\
    &=\frac{1}{2} \ex{\{\hat \phi(t),\hat \phi(t')\}},
\end{align}
where we have expanded the anti-time and time orderings in terms of theta functions and used $\theta(x)+\theta(-x)=1$. At last, a similar calculation gives
\begin{align}
    \exx{\varphi_{a}(t)\varphi_{a}(t')}&= \exx{\left[ \varphi_{+}(t)-\varphi_{-}(t) \right] \left[ \varphi_{+}(t')-\varphi_{-}(t') \right]}\\
    &=\ex{\T \hat \phi(t)\hat \phi(t')}+\ex{\bar\T \hat \phi(t)\hat \phi(t')}-\ex{  \hat \phi(t)\hat \phi(t')}-\ex{  \hat \phi(t')\hat \phi(t')}=0\,.
\end{align}
This illustrates the fact that $\varphi_{a}(t)$ cannot be considered as a propagating degree of freedom.

 
\section{Dissipative free theories}\label{sec:scalars}

This section aims at providing a general discussion of how to count the number of degrees of freedom for general dissipative but free theories. 
We first show that a theory is unitary if and only if it has a Hermitian Hessian matrix. This leads to two distinct ways in which a theory can be dissipative: when respecting time reversal or violating it. We point out that the quadratic Hessian matrix may be non-diagonalizable, but can always be put into Jordan form and this gives a canonical solution to the dynamics of the system. At last, we discuss the number of propagating degrees of freedom in constrained systems and in the presence of gauge symmetries.


\subsection{Dissipation and non-Hermitian Hessian matrices} 

To keep the discussion general, we consider an open system made of $n$ scalar fields 
\begin{align} 
S 
&=\int d^4x \pa^i(x) M_{ij}(\partial_t,\partial_i^2) \pr^j(x) =\int \frac{d^4k}{(2\pi)^4} \pa^{i}(-\omega,-\bfk) M_{ij}(\omega,k^2) \pr^{j}(\omega,\bfk)\,,\label{eq:Mdef}
\end{align}
and introduce the notation
\begin{align}
    S &=\int \pa^i M_{ij} \pr^j,
\end{align}
where $  \pa $ and $  \pr $ are  vectors with $  n $ components representing the advanced and retarded fields respectively and $  M $ is an $  n\times n $ matrix of differential operators acting on $\pr$. The matrix $ M$ is neither Hermitian nor anti-Hermitian, but must obey the non-equilibrium constraints discussed in \Sec{subsec:NEQconst}. Without loss of generality, it can be decomposed as
\begin{align}
M^{H}&=\frac{1}{2}\left(  M+M^{\dagger}\right)\,, & M^{A}&=\frac{1}{2}\left(  M-M^{\dagger}\right)\,, \then M=M^{H}+M^{A}\,.
\end{align}
The Hermitian conjugate operation can be understood equivalently in position or Fourier space. In Fourier space, we have $  \partial_{\mu}\sim  ik_{\mu} $ and so $  M $ can be truly thought of as a matrix. Then, Hermitian conjugate is simply the complex conjugate transposed. In position space, everything is real but we have to deal with derivatives and to specify where they are acting. The Hermitian conjugate requires integrating all derivatives by part, which brings about the same number minus signs as in Fourier space, namely one per derivative. As a simple example consider a single field with an interaction $  \pa \dot{\phi}_r  $. Here, the $1\times1$ matrix $  M\sim i \omega $ is not Hermitian but anti-Hermitian.

The non-equilibrium constraint \eqref{eq:herm} implies in Fourier space\footnote{In position space, the statement simply reads
\begin{align}
    M_{ij}=M_{ij}^*\in\mathbb{R} \then M^H_{ij}=M^H_{ji}\,, \quad  M^A_{ij}= - M^A_{ji}\,.
\end{align}} 
\begin{align}
    M_{ij}(k^\mu)=M_{ij}^*(-k^\mu) \then M^H_{ij}(k^\mu)=M^H_{ji}(-k^\mu) \,, \quad  M^A_{ij}(k^\mu)=-M^A_{ji}(k^\mu) \,.
\end{align}
Then, we can easily return to the $ \pm$ basis and write the open functional as
\begin{align}\label{eq:Msplit}
S=\frac{1}{2} \int \phi_{+}M^{H} \phi_{+} - \phi_{-} M^{H} \phi_{-}+ 2 \phi_{+}M^{A} \phi_{-} \,,
\end{align}
where we used the symmetry and anti-symmetry properties to simplify this expression. From this, we see that \textit{the theory is unitary if and only if $M$ is Hermitian}. Indeed, only when $M^{A}$ vanishes the theory reduces to two copies, one for each branch of the path integral.


\paragraph{Dissipation in time-reversal symmetric systems} Notice that at this point, there is no obstacle to write a dissipative theory that is also Lorentz invariant. For example, simply choose some non-zero $ M^{A}$ with numerical entries. As a minimal example, consider
\begin{align}
M=\begin{pmatrix} \omega^{2}-k^{2} & a \\ -a & \omega^{2}-k^{2} \end{pmatrix}\,.\label{eq:Mdeom}
\end{align}
This example is anti-Hermitian and real, so also anti-symmetric. The associated equations of motion are
\begin{align}\label{deom}
-\ddot \phi + \partial_{i}^{2}\phi +a\chi&=0\,, & -\ddot \chi + \partial_{i}^{2}\chi -a\phi&=0\,,
\end{align}
where we denoted $\phi_1 = \phi$ and $\phi_2 = \chi$.
This does not look particularly strange, but the different sign in the term proportional to $  a $ in the two equations leads to a non-conservation of energy. To see this, consider the unitary theory
\begin{align}
S_{\mathrm{unit}}=\int \dd^4x \left( -\frac{1}{2}\partial \phi^{2}-\frac{1}{2}\partial \chi^{2}+a\chi \phi\right)\,.
\end{align}
This has energy
\begin{align}
T_{00}&=\frac{1}{2}\left(  \dot \phi^{2}+\dot\chi^{2}+\partial_{i}\phi^{2}+\partial_{i}\chi^{2}-a\phi\chi\right)\,,
\end{align}
and equations of motion
\begin{align}\label{ueom}
-\ddot \phi + \partial_{i}^{2}\phi +a\chi&=0\,, & -\ddot \chi + \partial_{i}^{2}\chi +a\phi&=0\,.
\end{align}
It is straightforward to see that that in this unitary theory $  \partial_{\mu}T^{\mu0}=0 $, i.e. energy is conserved on the solutions of the equations of motion \eqref{ueom}. If instead we use the dissipative equations of motion in \eqref{deom}, we find that $  T^{\mu0} $ is not conserved because of the crucial different sign in the term proportional to $  a $. In this latter case, there is a non-reciprocal transfer between the two scalars, and so the name of \textit{non-reciprocity} used to characterise this dissipative setting. Because this theory is invariant under time reversal, the dispersion relation is a real equation and the frequencies that solve is come in complex conjugate pairs, $\omega_1$ and $\omega_2=\omega_1^*$. In the presence of dissipation these frequencies become complex and hence one of the two frequencies has an imaginary part whose sign leads to an exponential instability $e^{\pm i \omega t}\propto e^{+|\Im\omega|t}$. It would be interesting so see if, despite this instability, these non-reciprocal behavior can be useful to describe some physical system, but we leave this for future work. 

\paragraph{Dissipation from the breaking of time reversal symmetry} The above example is time-reversal symmetric. A different way in which a theory can be dissipative is by breaking time reversal. The prototypical example is an interaction with a single time derivative, as in the single scalar example discussed in \Sec{subsec:dispreleg} we reproduce here for clarity
\begin{align}
S=\int  \dd^4 x \left(  \dot{\phi}_a\dot{\phi}_r - \partial_{i}\pa \partial^{i}\pr + \pa \dot{\phi}_r  \right) \sim \int \pa(\omega^{2}-k^{2}+i\omega)\pr\,.\label{eq:Tbreaking}
\end{align}
The corresponding Hessian matrix is $  1\times 1 $ complex and so fails to be Hermitian. In contrast to the non-reciprocal case above, now the dispersion relation is not real and so its solutions do not come in complex conjugate pairs. Instead, given a solution $\omega_1$ we find a second by $\omega_2=-\omega_1^*$. This second solution $\omega_2 $ may or may not be the same as $\omega_1$. When $\Re\omega_1\neq 0$, then necessarily $\omega_2 \neq \omega_1$. For simple cases where $M$ is at most quadratic in $\omega$, if one solution has the correct sign of the imaginary part to guarantee stability, so has the second solution. In this concrete example one finds \Eq{eq:soldispeg}, that is
\begin{align}
   \omega_{1} = -i\frac{\gamma}{2}+ \sqrt{ k^{2}-(\gamma/2)^{2}},\quad \mathrm{and} \quad\omega_{2} = -i\frac{\gamma}{2}- \sqrt{ k^{2}-(\gamma/2)^{2}}.
\end{align}
It is easy to check explicitly that the imaginary parts of $\omega_1$ and $\omega_2$ indeed share the same sign, which entails the late-time stability of the system. 

 
\subsection{Number of degrees of freedom}\label{ssec:Jordan}

In this section, we derive a formula to compute the number of propagating degrees of freedom for a quadratic dissipative theory. When the Hessian matrix $M$ is diagonalizable, the final formula is given in \Eq{mex}. However, in contrast to what one finds for a unitary theory, in the dissipative case there is a set of theories for which the Hessian matrix is not diagonalizable. In this finely tuned special case the number of degrees of freedom is computed according to \Eq{dofgeneral}.\\

\paragraph{Diagonalizable case} We begin by going to Fourier space so that the Hessian matrix $  M $ has components that are simply functions of $  \omega $ and $  \bmk $. The equations of motion are simply given by the vector equation $M(\omega,k^2) \phi_r=0$. When this relation is satisfied, $\phi_r$ must be in the kernel of $M$ by definition. One would then tentatively propose that the number of degrees of freedom is given by
\begin{align}\label{toonaive}
\#_{\text{d.o.f.}}\overset{??}{=}\text{max}_{\omega}\, \text{Nul}(M) \quad \text{(naive)} \,,
\end{align}
where the nullity Nul is the dimension of the kernel of the linear operator $  M $. In a vector space of dimension $n$ it satisfies $ 0\leq \text{Nul} \leq n  $. This is too naive because there can be different $  \omega $'s and $  \bmk $'s for different degrees of freedom if they don't talk to each other. For example, in a simple toy model of two decoupled fields we have
\begin{align}\label{2dof}
M=\begin{pmatrix} \omega_{1}^{2}-k_{1}^{2}-m_{1}^{2} & 0 \\ 0 & \omega_{2}^{2}-k_{2}^{2}-m_{2}^{2} \,, \end{pmatrix}
\end{align} 
with two different $  \omega $'s. The vanishing of the two eigenvalues gives us the two dispersion relations $  \omega_{i}^{2}=k_{i}^{2}+m_{i}^{2} $ for the two fields $  i=1,2 $. So to get the right result, we have to perform the maximization in \eqref{toonaive} over two different frequencies $  \omega_{1,2} $. To make this precise, we have to specify what $\omega$'s in $M$ can be different. This is straighforward if $M $ is diagonalizable. Indeed, in the diagonal basis every eigenvector evolves independently of the others. The equation of motion for each eigenmode is simply given by the vanishing of the corresponding eigenvalue, which is nothing but the dispersion relation of that eigenmode. Each eigenmode can have a different $\omega$ because it does not mix with the others. The number of degrees of freedom is then just the number of eigenvectors that can be made to vanish by appropriately choosing a set of $n$ frequencies $\omega_i$, one for each eigenmode. This number is just the nullity of $M$, which of course does not depend on the basis in which it is computed. We are hence led to improve our previous relation to 
\begin{align}\label{mex}
\boxed{\#_{\text{d.o.f.}} =\text{max}_{\{\omega_{i}\}}\, \text{Nul}(M) \qquad \text{(diagonalizable)}}\,,
\end{align}
where $  i=1,\dots,n $ are the frequencies of each eigenmode.

If we were discussing unitary theories for which $M$ is Hermitian, this would be the end of the story because every Hermitian matrix is diagonalizable. However, as we have seen, in dissipative theories $M$ is not Hermitian and diagonalizability is not guaranteed. In particular, there is a set of non-diagonalizable Hessian matrices of measure sure. Non-diagonalizability is rare and requires fine tuning since diagonalizable matrices are dense in the space of matrices. We turn this special case next.


\paragraph{Non diagonalizable matrices and Jordan form} To understand the problem, let us look at the properties of $  M $ more closely. As we discussed, in the general dissipative theory, $M$ is not Hermitian, and hence it is not necessarily diagonalizable. This is a new phenomenon of the dissipative theory because in the unitary theory $M$ has to be Hermitian as proven above. As a canonical example consider
\begin{align}\label{toy2}
S=\int (\chi_{a},\sigma_{a}) \begin{pmatrix} \omega^{2}-m_{\chi}^{2} & -i \omega \\ 0& \omega^{2}-m_{\sigma}^{2} \end{pmatrix} \begin{pmatrix} \chi_{r} \\ \sigma_{r} \end{pmatrix}\,.
\end{align}
The non-unitary mixing $  -i\omega \chi_{a}\sigma_{r} $ makes $  M $ non-diagonalizable. This means there are not two eigenmodes but less. In this case the equations of motion are
\begin{align}
\ddot \chi_{r}+m_{\chi}^{2}\chi_{r}&=-\dot \sigma_{r} \,, & \ddot \sigma_{r}+m^2_{\sigma}\sigma_{r}=0\,.
\end{align} 
One eigenmode corresponds to $  \sigma =0 $ and $  \chi $ an harmonic oscillator with frequency $  m_{\chi} $. Put simply, the solution is simply an harmonic oscillation for $  \sigma $ which is then a source for $  \chi $. As long as they are not at the exact same frequency (or if there is dissipation), the solution is stable at late times. For example, the classic steady state solution of the forced harmonic oscillator ($  \chi $ in this case) is oscillating at the same frequency as the force ($  \sigma $ in this case) with a possible phase shift in the case of dissipation. As far as we can tell, there is nothing wrong with this system. It just does not have as many eigenmodes as one would expect, but has as many degrees of freedom. What special about is that two solutions do not exist that remain each itself as time progresses.

Loosely speaking, the example in \eqref{toy2} covers all possible ways in which $  M $ fails to be diagonalizable. Indeed, although not all matrices can be diagonalized, all matrices can be ``almost'' diagonalised (in an algebraically closed field such as the complex numbers) if we put then into Jordan (normal) form. In Jordan form a matrix is block-diagonal and every block has a single eigenvalue on the diagonal and the number one on the superdiagonal. There can be, and in general are, blocks with the same eigenvalue. A diagonal matrix is one for which the Jordan form has only $  1\times 1 $ blocks and is therefore diagonal. In fact, the number of blocks corresponding to a given eigenvalue is the dimension of the subspace spanned by the eigenvectors of this eigenvalue, which is called the ``geometric'' multiplicity of the eigenvalue. The number of times an eigenvalue appears in any block\footnote{Equivalently the sum of the size of all Jordan blocks with that eigenvalue.} is the ``algebraic multiplicity'' of that eigenvalue, simply coming from solving the characteristic polynomial. From this we see that a matrix is diagonalizable if and only if the geometric multiplicity is the same as the algebraic multiplicity (in general the former can be smaller than the latter, but never bigger). \\

Given a generic $  M $ in \eqref{eq:Mdef}, let us consider its Jordan form. Let $  1\leq J \leq n$ be the total number of Jordan blocks and let $  \lambda_{p} $ be the corresponding eigenvalues with $  p=1,\dots,J $. The vectors corresponding to a Jordan chain are those that span the subspace where the Jordan block acts. These vectors are coupled together as in the examples above, see \eqref{toy2}. As a consequence, all these vectors must have the same frequency $  \omega_{p} $, but different blocks may have different frequencies $  \omega_{p}\neq \omega_{p'}$. 
This finally  allows us to formulate our final equation for the number of degrees of freedom, which is valid both for diagonalizable and non-diagonalizable matrices:
\begin{align}\label{dofgeneral}
\#_{\text{d.o.f.}}=\text{max}_{\{\omega_{p}\}}\, \text{Nul}(M) \qquad \text{(general case)}
\end{align}
where again $  p $ runs over all the Jordan blocks. This formulae correctly predicts that both \eqref{2dof} and \eqref{toy2} describe theories with two degrees of freedom. It also tells us that \eqref{2dof} has two dispersion relations, corresponding to the vanishing of its two distinct eigenvalues, while \eqref{toy2} has a single dispersion relation, corresponding to the vanishing of its single eigenvalue. Of course \eqref{dofgeneral} reduces to \eqref{mex} when $M$ is diagonalizable and all Jordan blocks are $1\times 1$.


\subsection{Constrained degrees of freedom}

The power of \Eq{mex} is appreciated when the number of degrees of freedom is less than expected. We consider three examples: the first and the second involving two scalar fields, with the second being a truncated version of General Relativity (GR), and the third being Proca theory. 

\paragraph{Example 1: One dynamical and one constrained scalar} Let's begin considering 
\begin{align}
M=\left(
\begin{array}{cc}
 \omega ^2-k^2 & k^2-\omega ^2 \lambda  \\
 -\omega ^2+k^2 \lambda  & \omega ^2-k^2 \\
\end{array}
\right)\,,
\end{align}
where $  \lambda \neq 1 $ is some coupling constant. It is not clear a priori what $ \#_{\text{d.o.f.}}  $ should be. To compute it, we first reduce $  M $ to Jordan form and we realize it is actually diagonalizable (as it should be since it is symmetric). Then, we consider the eigenvalues $  (1-\lambda) k^{2} $ and $  \omega^{2}-(\lambda+1)k^{2} $. We have two eigenvectors and so we can vary the frequency $  \omega $ in each eigenvalue independently. We can choose $  \omega^{2}=(\lambda+1)k^{2} $ to make the second eigenvalue zero. However, the first eigenvalue does not depend on $  \omega $, and, for any non-zero momentum, it does not vanish. Hence, the maximum the nullity can be is one and so this theory describes one propagating degree of freedom. The simple change of variables that diagonalizes $  M $ tells us that one linear combination $  \phi_{r} $ of the fields obeys $  \nabla^{2}\phi_{r}=0  $ and is hence a constrained field. The avatar of having a constrained field is that $  \lambda(\omega_{p})=0 $ has no solution. In this example, this happened because $ \lambda$ did not depend at all on $ \omega$ but this phenomenon can happen more generally.

\paragraph{Example 2: Two constrained fields} Here, we consider the example of two scalar fields, which actually arises as a truncation of general relativity to two scalar perturbations. As before, we will find that $ \lambda(\omega)=0$ has not solution, but the reason now is different. We will see that the eigenvalues $  \lambda $ do depend on $ \omega$, but they are not polynomial functions of $  \omega $ and so $  \lambda(\omega_{p})=0 $ need not have a solution. The concrete example, is
\begin{align}
M=\left(
\begin{array}{cc}
 k^2-\omega ^2 &  f(k^2)   \\
g(k^2)   & 0 \\
\end{array}
\right)\,,
\end{align}
where $ f(k^{2})$ and $ g(k^{2})$ are local functions of the norm of the spatial vector $ k^{i}$ and can be both complex and different from each other, hence encompassing dissipative phenomena.  In GR, which is a non-dissipative theory, this happens with $ f(k^{2})=g(k^{2})=-k^{2}$. The matrix is diagonalizable for generic functions $ f$ and $ g$. The eigenvalues are found to be 
\begin{align}
\lambda_{1,2}= \frac{1}{2} \left(k^2-\omega ^2\pm\sqrt{fg +(k^{2}- \omega ^2)^{2}}\right) \,.
\end{align}
These never vanish for non-zero $  k $. To see this assume $ \lambda_{1,2}=0$, isolate the square root and square each side of the equation to discover $ 4k^{4}=0$. The above rule would conclude that this system does not propagate any degree of freedom. We can recover this result by looking directly at the equations of motion $ M\phi=0$, that is
\begin{align}
\ddot \phi -\partial_{i}^{2}\phi+\partial_{i}^{2}\chi&=0\,, & \partial_{i}^{2}\phi&=0\,.
\end{align}
The second equation tells us that $ \phi$ is a constrained field and does not propagate. Indeed the only solution that vanishes at spatial infinity is $ \phi=0$. Using this result in the first equation we get $ \partial_{i}^{2}\chi=0$, which tells us that also $ \chi $ is constrained and in fact must vanish everywhere.


\paragraph{Example 3: Proca theory} As a last example of constrained fields, we consider the theory of a massive vector field, a.k.a. Proca theory. Since our focus is on the counting of degrees of freedom rather than on modelling generic dissipative phenomena, we focus here on the unitary case. Our interest in this model is that it has a constrained field, but there is no gauge symmetry. This further elucidates how to deal with constraints before we move on to gauge theories in the next section. 
We want to describe the theory in terms of a four-vector field $A^\mu$. Since a massive vector propagates three degrees of freedom we will encounter a constraint. We work in the Keldysh basis by introducing
\begin{align}
A^{\mu}&=\frac{1}{2}\left( A_{+}^{\mu}+A_{-}^{\mu}\right) \,, & 
a^{\mu}=A_{+}^{\mu}-A_{-}^{\mu}\,.
\end{align}
We will restrict ourselves to the free theory. We take the open functional to be
\begin{align}
S 
=\int a^{\mu} M_{\mu\nu} A^{\nu}= \int a^{\mu} \left[  \eta_{\mu\nu}(\omega^2-k^2-m^2)+k_{\mu}k_{\nu}\right] A^{\nu}\,.
\end{align}
where $m$ is a mass and remind the reader that our signature is $(-,+,+,+)$ and in our notation $k^2\equiv |\bmk|^2$. When $m=0$, we have a theory with gauge invariance. In that case we know $S$ remains invariant under the shift $A_\mu\to A_\mu+\partial_\mu \epsilon$ and hence it must be that $ \text{det}M=0 $ for generic values of $  \omega $ and $  \bmk $. The mass term breaks gauge invariance and indeed we find
\begin{align}
    \det M =m^2 (\omega^2-k^2-m^2)^3\,,
\end{align}
which is non zero but vanishes for $m^2=0$. The eigenvalues of $M$ are
\begin{align}
    \left\{\omega^2 -k^2 - m^2 ,\, \omega^2 -k^2 - m^2 ,\, 
 \frac12 \left(k^2 + \omega^2 \pm 
    \sqrt{(k^2 + 2 m^2)^2 + 2 (k^2 - 2 m^2) \omega^2 + \omega^4} \right) \right\}\,.
\end{align}
According to our general formula, to count the number of degrees of freedom we have to ask how many of these dispersion relations can be satisfied by some choice of $\omega$, allowing for the fact that a different $\omega$ can be chosen in each eigenvalue. Clearly, the fist two eigenvalues vanish when $\omega^2=k^2+m^2$. Concerning the last two eigenvalues, demanding that they vanish and squaring the square root gives again $\omega^2=k^2+m^2$, which sets one of the two to zero but not the other. We conclude that only 3 eigenvalues can vanish and this is therefore the number of propagating degrees of freedom, as expected for a Proca field.

 
\section{Open Maxwell theory}\label{sec:OEM}

We now construct an open version of electromagnetism, namely a dissipative theory for a massless spin $1$ photon. To be precise, the first step in the construction of any EFT is to specify the number of degrees of freedom. Here we want to consider two degrees of freedom with helicity $\pm 1$. These describe the propagation of light in a typical dielectric material, a.k.a. an insulator. Many other interesting phenomena associated with the propagation of light in a medium are not captured by this theory. For example, in a conductor, the mobile charges present in the medium provide an additional degree of freedom and the end result is that the photon acquires a mass and the low energy dynamics contains \textit{longitudinal} plasma oscillations as well as transverse oscillations. The open EFT of such a system would require additional degrees of freedom that not included in the Open EFT we study in this work.

Our end result is summarised in the following.

\begin{tcolorbox}[%
			enhanced, 
			breakable,
			skin first=enhanced,
			skin middle=enhanced,
			skin last=enhanced,
			before upper={\parindent15pt},
			]{}

\vspace{0.05in}

\paragraph{Summary on open E\&M:} To all order in derivatives and in the free theory, the open effective functional $S= S_1 + S_2$ of open E\&M is
\begin{align}
S_1=\int_{\omega,\bmk}\left[ a^{0}  i k_{i}F^{0i}+a_{i}\left(  \gamma_{2}F^{0i}+\gamma_{3}ik_{j}F^{ij}+\gamma_{4}\e^{i}_{\,jl}F^{jl}\right)\right]\, \equiv \int_{\omega,\bmk}  a^{\mu}M_{\mu\nu}A^{\nu}\,,
\end{align}
and 
\begin{equation}
S_2= i \int_{\omega,\bmk} a^{\mu}N_{\mu\nu}a^{\nu},
\end{equation}
where $N_{\mu\nu}$ is any $4\times4$ positive definite matrix and $\gamma_{2,3,4}$ are model-dependent analytic functions of $\omega$ and $k^2$. $S_1$ remains unchanged under the following retarded and advanced gauge transformations
\begin{align}
A^\mu &\rightarrow A^\mu + \e_r k^\mu,& a^\mu \rightarrow a^\mu + \e_a v^\mu,
\end{align}
where $k^\mu = (\omega,\bmk)$ and $ v^\mu = (i\gamma_2, \bmk)$. The equations of motion in the presence of external sources $j^\mu$ read $M_{\mu\nu}A^{\nu}= j_{\mu}+\xi_{\mu}$ where $\xi_\mu$ are stochastic noises that obey the constraint
\begin{align}
    v^{\mu}(j_{\mu}+\xi_{\mu})=0.
\end{align}
At last, the dispersion relations for the two helicities are given by 
\begin{align}
    i \gamma_{2} \omega+ \gamma_{3}  k^2 \pm 2 \gamma_{4} k  = 0,
\end{align}
from which we recover the results of E\&M in a medium in \Sec{subsec:standardEM}.
\end{tcolorbox}


\subsection{Effective functional}\label{subsec:functional}

We begin by doubling the gauge fields, $A^\mu \to A^\mu_\pm$, and expressing the result in the Keldysh basis
\begin{align}
\text{retarded: } A^{\mu}&=\frac{1}{2}\left( A_{+}^{\mu}+A_{-}^{\mu}\right) \,, & 
 \text{advanced: } a^{\mu}=A_{+}^{\mu}-A_{-}^{\mu}\,.
\end{align}
In the dissipative theory, there is an unbroken gauge transformation acting on both $A_+$ and $A_-$ corresponding to direction $ \e_{+}=\e_{-}=\e$. Under this gauge transformation $ A^{\mu}$ transforms as the usual gauge field, while $ a^{\mu}$ does not transform:
\begin{align}\label{eq:gaugetrans}
\text{retarded gauge transformation: } A^{\mu}&\to A^{\mu}+\partial^{\mu}\e\,, & a^{\mu}&\to a^{\mu}\,.
\end{align}
A preliminary analysis of the number of degrees of freedom proceeds as follows. The advanced fields $ a^{\mu}$ must satisfy a boundary condition problem, being constrained to vanish at initial and final time, so they cannot propagate. Of the four retarded degrees of freedom $ A^{\mu}$, we would like only two to propagate. We expect this will happen by gauge fixing one of them and having a second one fixed by a constraint equation. Both of these facts require the action to be gauge invariant (otherwise we cannot ``gauge fix'' and in fact the would-be ``constraint equation'' would become dynamical). To write down a gauge invariant effective functional, \Eq{eq:gaugetrans} indicates that we can use $ a^{\mu}$ at will, since it is gauge invariant. On the other hand, we should use the retarded $A^\mu$ fields in the (retarded) field strength combination
\begin{align}
F^{\mu\nu}=\partial^{\mu}A^{\nu}-\partial^{\nu}A^{\mu}\,.
\end{align}
Therefore the building blocks of our construction will be $F^{\mu\nu}$ and $a^\mu$. One can wonder whether this construction exhausts the set of allowed operators. In \Sec{subsec:SantiEM}, we prove that this is indeed the case by solely considering the gauge invariant quadratic forms one can write in the Schwinger-Keldysh contour. This guarantees the completeness of the operator basis.

Since E\&M is a free theory, we can now write all possible quadratic terms, starting at least linear in the advanced fields to fulfill the non-equilibrium constraints \eqref{eq:NEQ}.
Importantly, we work directly to all orders in derivatives. Indeed, to linear order (quadratic action) in flat space, we can formally specify an open functional in Fourier space that permits such construction. We show in \Sec{subsec:standardEM} how to make contact with the way E\&M in a medium is usually discussed. 

Let us first consider the terms linear in the advanced fields $ a^{0}$ and $ a^{i}$, which control the dispersion relations of the propagating degrees of freedom \cite{kamenev_2011, Salcedo:2024smn}. The most general open functional reads
\begin{align}
S_1 = \int \frac{\dd \omega}{2\pi} \int \frac{\dd^3 \bmk}{(2\pi)^3} \left[a^{0}(\gamma_{ij}^{t}F^{ij}+\gamma^{t}_{i}F^{0i}+j_{0})+a^{i}\left(  \gamma^{s}_{ij}F^{0j}+\gamma^{s}_{ijl}F^{jl}+j_{i}\right)\,\right].
\end{align}
where  ``$t $'' stands for time and ``$s$'' for space.
Assuming homogeneity and isotropy, we can write the EFT coefficients as
\begin{align}
\gamma^{t}_{i}&=\gamma^{t}(\omega,k) ik_{i}\,, & \gamma^{t}_{ij}&=\gamma^{t}_{2}(\omega,k)k_{i}k_{j}+\gamma^{t}_{3}(\omega,k)\delta_{ij}\,, \\
\gamma^{s}_{ij}&=\gamma_{1}(\omega,k)k_{i}k_{j}+\gamma_{2}(\omega,k)\delta_{ij}\,, & \gamma_{ijl}^{s}&=\gamma_{3}(\omega,k)\delta_{i[j}ik_{l]}+\gamma_{4}(\omega,k)\e^{ijl}\,.
\end{align}
where we introduced seven scalar functions of frequency and momentum. Also, we inserted an $i$ for every derivative for later convenience. 
The non-equilibrium constraints \eqref{eq:norm}, \eqref{eq:herm} and \eqref{eq:pos} impose a reality condition on the EFT coefficient of $S_1$ in real-space. This translates into conditions on the frequency-space coefficients, for instance \cite{Crossley:2015evo}
\begin{align}
    \gamma_2(\omega, k) = \Gamma - i \omega + \Gamma_{20}  \omega^2 + \Gamma_{02}  k^2 + i \Gamma_{30} \omega^3 + \cdots
\end{align}
that we normalized $\gamma_2(\omega, k) $ in this way for later convenience. It readily follows that the non-equilibrium constraints guarantee the $\Gamma$'s coefficients are all real.

Notice that one cannot construct an anti-symmetric two-index tensor and so $ \gamma^{t}_{ij}F^{ij}=0$. Also, notice that $ \gamma_{ijl}^{s}$ had to be anti-symmetric in $ jl$. As a consequence, we are left with
\begin{align}\label{OEM}
S_1&= \int \frac{\dd \omega}{2\pi} \int \frac{\dd^3 \bmk}{(2\pi)^3} \left[ a^{0}(\gamma^{t} i k_{i}F^{0i}+j_{0}) +a_{i}\left( \gamma_{1}k^{i}k_{j}F^{0j}+ \gamma_{2}F^{0i}+\gamma_{3}ik_{j}F^{ij}+\gamma_{4}\e^{i}_{\,jl}F^{jl}+j_{i}\right)\right]\,. \Bigg.
\end{align}
A few coefficients can be removed by rescalings and field redefinitions. First, one can set $\gamma^{t}=1$ (or to any other value) by rescaling $a^{0}$. Then, the term proportional to $\gamma_{1}$ can be removed by the field redefinition
\begin{align}
a^{0}\to a^{0}-i\gamma_{1}k_{i}a^{i}\,.
\end{align}
It is a simple exercise to keep $\gamma^{t}$ and $\gamma_{1}$ to confirm they do not have any consequence for the dynamics. In the following, we simply set $\gamma_{t}=1$ and $ \gamma_{1}=0$. 
Moreover, we momentarily neglect the currents, as these can be easily re-inserted at the end of the calculation. As a result, we land on the following theory
\begin{align}\label{OEMnog}
S_1=\int \frac{\dd \omega}{2\pi} \int \frac{\dd^3 \bmk}{(2\pi)^3} \left[ a^{0}  i k_{i}F^{0i}+a_{i}\left(  \gamma_{2}F^{0i}+\gamma_{3}ik_{j}F^{ij}+\gamma_{4}\e^{i}_{\,jl}F^{jl}\right)\right]\,.
\end{align}

\subsection{Gauge invariance and gauge fixing}\label{subsec:gauge}

We want to understand the dynamics of this theory in terms of the general formalism for free dissipative theories discussed above. To this end, we re-write the above functional in the form
\begin{align}\label{openf}
S_1=\int \frac{\dd \omega}{2\pi} \int \frac{\dd^3 \bmk}{(2\pi)^3}   a^{\mu}M_{\mu\nu}A^{\nu}\,,
\end{align}
with 
\begin{align}\label{eq:matrix1}
M=\begin{pmatrix}  k^{2} & - \omega k_{i} \\ -i \gamma_{2} k_{i} & i\gamma_{2}\omega \delta_{ij}+\gamma_{3}(k^{2}\delta_{ij}-k_{i}k_{j})-2i\gamma_{4}\e_{ijl}k_{l} \end{pmatrix}\,.
\end{align}
A few comments apply. First, it is straightforward to check that this matrix has a vanishing determinant, as it should be since the gauge invariance \eqref{eq:gaugetrans} must imply $ M_{\mu\nu} k^{\nu}=0$ where $ k^{\mu}=(\omega,\mathbf{k})$. 
Importantly, the presence of a ``right kernel" due to gauge invariance implies the existence of a ``left kernel" $v^\mu M_{\mu \nu} = 0$.\footnote{Formally, instead of left and right kernels, we should rather discuss the null spaces of $M$ and $M^\dag$. The theory being formulated in Fourier space, the matrices $M$ and $M^\dag$ are in the end mapped into linear differential operators in real space. A well-posed problem necessitates boundary conditions, in which case the use of $M^\dag$ is probably more appropriate for a rigorous formulation. In practice, this amounts to use $w^\mu = (v^\mu)^*$  instead of $v^\mu$ which is inconsequential for the current discussion.} 
In general, the left kernel and the right kernel are spanned by different vectors since $M$ is non-Hermitian. For instance, an inspection of $M$ leads to the left kernel being spanned by 
\begin{align}
    v^\mu = (i\gamma_2, \bmk).\label{eq:vmu}
\end{align}
These observations are striking. Even if we did not impose any structure on the appearance of the advanced field $a^\mu$, the gauge invariance of the retarded sector enforces enough constraints to generate an \textit{advanced gauge invariance} in the advanced sector. 
\begin{tcolorbox}[%
			enhanced, 
			breakable,
			skin first=enhanced,
			skin middle=enhanced,
			skin last=enhanced,
			before upper={\parindent15pt},
			]{}

\vspace{0.05in}

\paragraph{Retarded and advanced gauge invariances:} $S_1$ remains unchanged under the transformations
\begin{align}\label{eq:advgauge}
A^\mu &\rightarrow A^\mu + \e_r k^\mu,& a^\mu \rightarrow a^\mu + \e_a v^\mu.
\end{align}
       
\end{tcolorbox}
\noindent As we will see shortly, this new symmetry extends to the noise sector as well and reduces the number of advanced components needed to describe the problem. Moreover, the difference between the left and right kernel is a reminder of the dissipative nature of the dynamics, since in the unitary case one would find $v^\mu=k^\mu$. This is to be expected because in the absence of dissipation one should recover the two independent gauge groups acting on each branch of the closed-time contour. It would be desirable to have an interpretation of the above transformation of $a^\mu$ in terms of differential geometry, just like we think of the standard gauge field as a connection on the principle fiber bundle. 

At last, notice that we reduce to the standard Maxwell theory by setting 
\begin{align}\label{Maxwell}
\text{Maxwell: }\gamma_{2}&=-i\omega\,, & \gamma_{3}&=-c^{2}=-1\,, & \gamma_{4}&=0\,,
\end{align} 
in which case we find that $M$ becomes Hermitian and, being also real, symmetric. This also means that the left kernel of $M$ is spanned by the same vector $ k^{\mu}$ as the right kernel. Again this shows that there are two copies of the standard E\&M gauge group in the absence of dissipation.

\paragraph{Gauge invariant properties} It is interesting to study the system before we commit to any specify gauge. The eigenvalues of $M$ are found to be
\begin{align}\label{eq:eigenM} 
(0, k^2 + i \gamma_2 \omega, 
k^2 \gamma_3 - 2 k \gamma_4 + i \gamma_2 \omega, 
k^2 \gamma_3 + 2 k \gamma_4 + i \gamma_2 \omega)\,.
\end{align}
The zero eigenvalue is a consequence of gauge invariance. The last two are seen to propagate healthy modes with the Maxwell value given above. Instead, the second eigenvalue in the Maxwell theory becomes $\omega^2+k^2$. Demanding that it vanishes to satisfy the classical equations of motion reveals a complex $\omega$ and so an instability. If we insisted in quantizing the theory this mode would lead to the familiar problem of negative norm states encountered when quantizing in Lorentz gauge. It would be interesting to develop the formalism further and find a consistent prescription for a gauge invariant quantization, which would probably borrow ideas from BRST \cite{Becchi:1974md,Becchi:1975nq,Tyutin:1975qk} and Fadeev Popov \cite{Faddeev:1967fc} for non-Abelian gauge theories. We leave this for future research. 

\paragraph{Gauge fixing} A popular option to study this system is to work in Coulomb gauge. In the unitary case, this makes manifest that there are only two propagating degrees of freedom, namely the transverse component of $ A^{\mu}$. Since fixing the gauge will also be very convenient in the case of gravity, here we see how gauge fixing works in the case of a dissipative theory. 

By a retarded gauge transformation, we can remove one component from the retarded field $A^{\mu}$. In Coulomb gauge, the condition is $\partial_{i}A^{i}=0$. Explicitly, this means 
\begin{align}
    \exists~\e_r \quad \mathrm{s.t.}\quad k_i A^{\prime i} = 0, \quad \mathrm{where} \quad A^{\prime \mu} = A^\mu + \e_r k^\mu.
\end{align}
One can easily solve to find
\begin{align}
    \e_r = -\frac{k_i A^i}{k_i k^i},
\end{align}
which illustrates how the gauge fixing solution crosses each gauge orbit once: for each value of $A_i$, there is a unique $\e_r$ such that $A^{\prime \mu}$ always fulfill the gauge fixing condition.

We may think from \Eq{eq:gaugetrans} that the initial gauge invariance of $a^{\mu}$ precludes the possibility to gauge fix any of its components. Yet, the important lesson of \Eq{eq:advgauge} is the emergence of a gauge invariance for the advanced sector which allows us to fix one component just as we did for the retarded sector. Among all possibilities, it turns out to be convenient to consider the \textit{advanced} Coulomb gauge, in which $\partial_i a^i = 0$. One can indeed reach this gauge from
\begin{align}\label{eq:advgf}
    \exists~\e_a \quad \mathrm{s.t.}\quad k_i a^{\prime i} = 0, \quad \mathrm{where} \quad a^{\prime \mu} = a^\mu + \e_a v^\mu.
\end{align}
Just as above, it admits solutions for
\begin{align}
    \e_a = - \frac{k_i a^i}{k_i v^i}.
\end{align}

\subsection{Dispersion relations}\label{subsec:disprel}

Let us consider the number of propagating degrees of freedom and their dispersion relation in the (retarded \textit{and} advanced) Coulomb gauges. Although this is not a necessary assumption, the physics becomes quite transparent assuming we are studying a plane wave propagating in the $\hat z$ direction. First, we impose the Coulomb gauges by projecting $A^{\mu}$ and $a^{\mu}$ on the space perpendicular to $\hat z$. It amounts to drop their last component. For a wave moving in an arbitrary direction this would boil down to dividing $A^{\mu}$ and $a^{\mu}$ into their longitudinal and transverse components,
\begin{align}
A^{i}=\Al^{i}+\At^{i}\,, \quad \text{with} \quad k_{i}\At^{i}=k^{j}\e_{ijl}\Al^{j}=0\,,
\end{align}
and similarly for $a^\mu$. Then we derive a $ 3\times 3$ linear operator that describes the open dynamics of the 3 components in $ A^{\mu}$ and $ a^{\mu}$ that are not fixed by a gauge condition by dropping the longitudinal directions.  Inserting $A^{i}=\At^{i}$ and $a^{i}=\at^{i}$ in the open functional in \eqref{openf}, one finds
\begin{align}
S= \int \frac{\dd \omega}{2\pi} \int \frac{\dd^3 \bmk}{(2\pi)^3}  \left[ k^{2} a^{0}A^{0} +\left(  i\gamma_{2}\omega+\gamma_{3}k^{2}\right)\at^{i}\At^{i}-2i\gamma_{4} \e_{ijl}k^{i}\at^{j}\At^{l}\, \right].
\end{align}
This can also be written as the linear operator
\begin{align}\label{eq:matrix2}
M_{\perp} =\begin{pmatrix}  k^{2} & 0 & 0  \\  0 &  i\gamma_{2}\omega+\gamma_{3}k^{2} &  -2i\gamma_{4} k\\ 0 &  2i\gamma_{4} k &  i\gamma_{2}\omega+\gamma_{3}k^{2}  \end{pmatrix}\,,
\end{align}
where the second and third rows and columns refer to the two independent components of $ \At$ and $ \at$, which are an orthonormal bases of the plane perpendicular to $ k^{i}$. The three eigenvalues are
\begin{align}\label{eq:disprelfull}
(k^2\,,  i \gamma_{2} \omega+ \gamma_{3}  k^2 + 2 \gamma_{4} k, i \gamma_{2} \omega+ \gamma_{3}  k^2 - 2 \gamma_{4} k )\,.
\end{align}
This result is consistent with our expectation of having two propagating degrees of freedom. First, we can see that there is no choice of $ \omega$ such that the $00$-component of $ M_{\perp} $ vanishes. This indicates the existence of at least one \textit{constrained} degree of freedom, which is not propagating. This tells us that $ A^{0}$ is a constrained field, as expected.\footnote{If one tries to derive dispersion relations without fixing the gauge, an analogous result can be reached. Among the three non-zero eigenvalues, two of them carry the propagating modes while the third one corresponds to a ghost dispersion relation for which $\omega$ always remains purely imaginary.} Second, we can diagonalize the remaining $2\times 2 $ block of $ M_\perp$ to find two eigenvalues. Demanding that they vanish gives us the dispersion relations for the two dynamical degrees of freedom (the two polarizations)
\begin{align}\label{eq:disprelfin}
 i \gamma_{2} \omega+ \gamma_{3}  k^2 \pm 2 \gamma_{4} k   =0\,.
\end{align}
To interpret this relation, let us first consider Maxwell theory obtained from \Eq{Maxwell}. In this case, the dispersion relation reduces to $\omega^{2}=c^{2}k^{2}$ as expected. This also tells us that the constant part of $ \gamma_{3}$ is (minus) the square of the speed of light in the medium, that is $ \gamma_{3}=-v^{2}$. As we will see in \Sec{subsec:standardEM}, it is easy to relate this parameter to the refractive index $n=1/v$. The other leading order effect is a constant real part in $ \gamma_{2}$, that is $\gamma_{2}(\omega,k^2) \simeq \Gamma -i\omega$, which introduces the standard dissipation
\begin{align}
\omega^{2} +i\Gamma \omega - v^{2}k^{2} = 0\then \omega=-i\frac{\Gamma}{2}\pm \sqrt{v^{2}k^{2}-(\Gamma/2)^{2}}\,.
\end{align}
For $ \Gamma >0$, this gives a stable system. At last, one can consider the effect of $\gamma_4$, leading to the dispersion relation
\begin{align}
\omega^{2} +i\Gamma \omega - v^{2}k^{2} \pm 2 \gamma_4 k = 0\then \omega=-i\frac{\Gamma}{2}\pm \sqrt{v^{2}k^{2}-(\Gamma/2)^{2} \mp 2 \gamma_4 k}\,.
\end{align}
The two polarizations now have distinctive propagations, an effect known as birefringence.\footnote{We observe an IR instability when $k \rightarrow 0$ for one of the two polarizations which may be cured by non-linearities. We leave its investigation for future works.} The discussion of the dispersion relations can be extended by accounting for higher-orders in derivatives. In general, $\gamma_{2,3,4}$ should be analytic functions of $ \omega$ and $ k^{2}$ around the origin by locality in time and space, assuming isotropy.


\subsection{Noise constraint}\label{subsec:noise}

The inclusion of noises requires the extension of the effective functional to quadratic operators in the advanced field \cite{kamenev_2011, Salcedo:2024smn}. In this section, we first present the Hubbard-Stratonovich trick which allows us to start from the open effective functional that includes such quadratic terms and obtain a linear functional in the advanced component. With the theory in that form, we can perform the path integral over the advanced component from which we derive the Langevin equation. Up to quadratic order in the advanced sector, the general open effective functional for open E\&M is
\begin{equation}
    S=\int d^{4}x \left[a^{\mu}M_{\mu\nu}A^{\nu}+ia^{\mu}N_{\mu\nu}a^{\nu}\right],
\end{equation}
where $M_{\mu\nu}$ and $N_{\mu\nu}$ contain differential operators and the $i$ comes from the non-equilibrium constraints.
The Hubbard-Stratonovich trick replaces the quadratic terms in $a^{\mu}$ through a path integral over an auxiliary field $\xi_{\mu}$
\begin{equation}
    \text{exp}\left[-\int d^{4}xa^{\mu}N_{\mu\nu}a^{\nu}\right]=\int[\mathcal{D}\xi_{\mu}]\;\text{exp}\left[-\int d^{4}x\left(\frac{1}{4}\xi_{\mu}(N^{-1})^{\mu\nu}\xi_{\nu}+ia^{\mu}\xi_{\mu}\right)\right].
\end{equation}
Note that the non-equilibrium constraints are fundamental in this step, as the convergence of the path integral requires that $N$ is positive-definite, and hence, invertible. In the end, the partition function of the theory is given by three path integrals:
\begin{equation}
    \mathcal{Z}=\int[\mathcal{D}A^{\mu}]\int[\mathcal{D}a^{\mu}]\int[\mathcal{D}\xi_{\mu}]\;\text{exp}\left[\int d^{4}x \;ia^{\mu}\left(M_{\mu\nu}A^{\nu}-j_{\mu}-\xi_{\mu}\right)-\frac{1}{4}\xi_{\mu}(N^{-1})^{\mu\nu}\xi_{\nu}\right]
\end{equation}
In this form, the advanced component behaves like a Lagrange multiplier and the constraints it enforces are the equations of motion known as the Langevin equation. 

Note that the classical external current $j_\mu$ appears together with $\xi_\mu$ in the term linear in $a^\mu$. Since we have chosen to define $\xi_\nu$ such that they have zero average, $\ex{\xi_\nu}=0$, the above observation suggests the interpretation of $j_\mu$ as the classical average of noise fluctuations. In other words, one could have just as well defined a new current $J_\nu\equiv j_\nu +\xi_\nu$ with a non-zero average, $\ex{J_\nu}=j_\nu$, but we will not use this notation in the following.

Importantly, the advanced gauge symmetry identified above leads to a decoupling of some directions from the noise and the currents, to which we will refer to as \textit{noise constraints}. We now illustrate how the advanced gauge symmetry $a^\mu \rightarrow a^\mu + \e_a v^\mu$ can be translated into the Langevin equation as a constraint on the stochastic noises and currents. First, we decompose $a^\mu$ in the longitudinal and transverse directions such that 
\begin{align}
    a^\mu = \lambda v^\mu + a^\mu_{\radot},
\end{align}
where $\lambda$ is a real proportionality coefficients and $\radot$ stands for the orthogonal direction to $v^\mu$. The integral over $a^{\mu}$ can be divided through this bipartition such that
\begin{equation}
\mathcal{Z}=\int[\mathcal{D}A^{\mu}]\int[\mathcal{D}\xi_{\mu}]\mathcal{Z}_{v}[\xi]\mathcal{Z}_{\radot}[\xi,A],
\end{equation}
where
\begin{align}
    &\mathcal{Z}_{v}[\xi]=\int\;d\lambda\;\text{exp}\left[\int d^{4}x \;i\lambda v^{\mu}\left(M_{\mu\nu}A^{\nu}-j_{\nu}-\xi_{\nu}\right)\right],\\
    &\mathcal{Z}_{\radot}[\xi,A]=\int[\mathcal{D}a_{\radot}^{\mu}]\;\text{exp}\left[\int d^{4}x \;ia_{\radot}^{\mu}\left(M_{\mu\nu}A^{\nu}-j_{\nu}-\xi_{\nu}\right)-\frac{1}{4}\xi_{\mu}(N^{-1})^{\mu\nu}\xi_{\nu}\right].
\end{align}
As $v^{\mu}$ belongs to the left kernel of $M$, there is no dependence on $A$ in $\mathcal{Z}_{v}[\xi]$. Performing the integral over $\lambda$ generates 
\begin{tcolorbox}[%
			enhanced, 
			breakable,
			skin first=enhanced,
			skin middle=enhanced,
			skin last=enhanced,
			before upper={\parindent15pt},
			]{}

\paragraph{Noise constraint:} 
\begin{equation}\label{eq:noiseconstraint}
    v^{\mu}(j_{\mu}+\xi_{\mu})=0.
\end{equation}       
\end{tcolorbox}

\noindent This is a constraint on the noise and the currents that arises as a result of the path integral along the advanced component parallel to the advanced gauge symmetry\footnote{Note that it is consistent with starting from the Langevin equation $M_{\mu\nu}A^{\nu} = j_{\mu}+\xi_{\mu}$ and taking the overlap with $v^\mu$ acting on the left, in which case one arrives at the same constraint equation for the noise and currents.}. We will re-derive this result in a different way around \eqref{eq:GBcondition}. To understand the noise constraint, we notice that for Maxwell's theory $v^\mu=(\omega,k^i)$ and so in the unitarity case this constraint simply enforces the conservation of the total current $\partial^\mu(j_\mu+\xi_\mu)=0$. In contrast, in the dissipative theory $v^\mu=(i\gamma_2,\bfk)$. Then, we can separate $\gamma_2=-i\omega+\Upgamma$ and write
\begin{align}
    ik^\mu (j_\mu+\xi_\mu)=\Upgamma (j_0+\xi_0)\,.
\end{align}
This tells us that the current that is perceived by the photon is \textit{not} conserved in the presence of dissipation. It is amusing to notice that this follows from advanced gauge invariance, $a^\mu \to a^\mu + \epsilon v^\mu$, just like the conservation of current follows from standard gauge invariance in the familiar non-dissipative spinor electrodynamics. 

Meanwhile, the integral over the orthogonal directions yield the reduced Langevin equation, where both sides are orthogonal to $v^\mu$
\begin{equation}
    \left(M_{\mu\nu}A^{\nu}\right)_{\radot}=(j_{\mu}+\xi_{\mu})_{\radot}.
\end{equation}
As one could have correctly suspected from the advanced gauge invariance, this illustrates how the four advanced fields play a degenerate role. Hence a linear combination of them can be removed without loss of generality. In the absence of currents, the two point function of the noise $\xi_{\mu}$ can be obtained from the projected inverse matrix of $N$
\begin{equation}\label{eq:noisePk}
    \langle\xi_{\mu}\xi_{\nu}\rangle=P_{\mu}^{\;\alpha}P_{\nu}^{\;\beta}(N^{-1})_{\alpha\beta}\,,
\end{equation}
where the projector $P=P^2$ removes the component along $v$, namely $v^\mu P_{\mu}^{\;\alpha}=0$.
In practice, this discussion amounts to removing one component of $\xi^\mu$, which in the absence of currents is simply parallel to $v^\mu$.


\subsection{A constructive approach}\label{subsec:SantiEM}

In this subsection, we provide an alternative and complementary perspective on the study of open E$\&$M. This \textit{constructive} approach focuses on the properties of spin 1 massless particles from which we build all possible EFT operators. It has the advantage of being extendable to spin 2 particles and justifies the uniqueness of some of the operators in the previous subsections. In this picture, the advanced gauge symmetry presents itself as a natural consequence of gauge invariance of the retarded sector and its relation to dissipation is straightforward. In the end, the connection between advanced gauge symmetry and the noise constraint arises naturally in this language. 
Despite a drastically different starting point, we arrive at the same conclusions, which is a great consistency check of our final results. Importantly, this approach is valid out-of-equilibrium and only relies on retarded gauge invariance, non-equilibrium constraints and the Lorentz breaking classification of operators.

\paragraph{Building operators} We start by considering the matrix $M$ in \Eq{eq:Mdef} to be a sum over operators
\begin{equation}\label{eq:matform}
     M_{\mu\nu}(\omega,\bmk)=\sum_{n}\mathcal{O}^{(n)}_{\mu\nu}(\omega,\bmk).
\end{equation}
We now construct and classify these operators, depending on whether they preserve Lorentz invariance or not, and whether they are unitary or not. Each of these operators is subject to different constraints and engenders different phenomena. For example, the simplest operator is Lorentz invariant and unitary. Then, we proceed with Lorentz violating unitary operators, which introduce a new time-like 4-vector $n^{\mu}$ associated with a preferred timelike direction that breaks Lorentz invariance.\footnote{Assuming the Lorentz group is broken to spatial homogeneity and isotropy, there cannot be more than one breaking direction. Indeed, if $n^\mu = (-1, \bm{0})$, a second non-trival $\tilde{n}^\mu = (\alpha^0, \bm{\alpha})$ must necessarily include a preferred spacial direction $\bm{\alpha}$, which further breaks the reduced symmetry group.} This construction leads to the identification of two operators whose lowest-order in derivative expressions can be related to a speed of sound and birefringence.
Once the discussion of these unitary operators is completed, we include the Lorentz violating and non-unitary operator whose lowest-order in derivatives expression relates to dissipation. We explain how its properties lead to the advanced gauge symmetry, distinct from that of the retarded sector. 


\paragraph{Lorentz invariant and unitary} The simplest operators to consider are both Lorentz invariant and unitary. Lorentz covariance implies that there are only two possible choices
\begin{equation}
    K_{\mu\nu}=\alpha_{1}(-\omega^2 + k^2)\eta_{\mu\nu}+\alpha_{2}k_{\mu}k_{\nu},
\end{equation}
where $\alpha_{1}$ and $\alpha_2$ are analytic functions of $ \omega$ and $ k^{2}$. Crucially, retarded gauge invariance $K_{\mu\nu} k^\nu = 0$ imposes
\begin{equation}
    \alpha_{1}+\alpha_{2}=0.
\end{equation}
An example to lowest order in derivatives is the kinetic operator. In this case, we canonically normalize the operator to be
\begin{equation}
    K_{\mu\nu}=(-\omega^2 + k^2)\eta_{\mu\nu}-k_{\mu}k_{\nu}.
\end{equation}


\paragraph{Lorentz violating and unitary} The next operators we consider are still unitary, but break Lorentz invariance to the Euclidean group SO(3), that is rotations and translations. Physically, they capture the fact that a homogeneous and isotropic material selects a preferred reference frame. An example at lowest order in derivatives is the well-known speed of sound operator. Mathematically, we account for the existence of a preferred reference frame by introducing a timelike direction $n^{\mu}$, which we normalize by $n^\mu n_\mu=-1$.\footnote{Bear in mind that $n^\mu$ is different from the advanced gauge direction $v^\mu$ identified above. The former relates to a preferred timelike direction while the later is the eigenvector spanning the null space of $M^{\mathrm{T}}$. Their relation is formally established below \Eq{eq:noiseconstraintSanti}.} The two-indices operators we can construct are
\begin{equation}
S_{\mu\nu}=\beta_{1}K_{\mu\nu}+\beta_{2} (-\omega^2 + k^2)n_{\mu}n_{\nu}+\beta_{3}\omega(k_{\mu}n_{\nu}+k_{\nu}n_{\mu})+\beta_{4}\omega^2\eta_{\mu\nu},
\end{equation}
where the $\omega$ and $\omega^2$ next to $\beta_3$ and $\beta_4$ comes from the contractions $\eta_{\mu\nu}k^\mu n^\nu$ and $(\eta_{\mu\nu}k^\mu n^\nu)^2$ respectively. Then, retarded gauge invariance $S_{\mu \nu} k^\nu = 0$ implies the last three terms combine into
\begin{equation}
S_{\mu\nu}=\beta_{1}K_{\mu\nu}+\beta_{2} [(-\omega^2 + k^2)n_{\mu}n_{\nu}-\omega(k_{\mu}n_{\nu}+n_{\mu}k_{\nu})+\omega^2\eta_{\mu\nu}],
\end{equation}
from which we still need to fix $\beta_{1}$ and $\beta_{2}$. Crucially, at lowest order in derivatives, the speed of sound operator does not involve any time derivative, such that it cannot couple to the derivative along the timelike direction $n^{\mu}$.
Therefore, we demand that the operator $S_{\mu\nu}$ vanishes for $k^\mu \propto n^\mu$. This requirement fixes the operator to be
\begin{equation}
S_{\mu\nu}=(c_{s}^{2}-1)[(-\omega^2 + k^2)n_{\mu}n_{\nu}-\omega(k_{\mu}n_{\nu}+n_{\mu}k_{\nu})+ k^2\eta_{\mu\nu}-k_{\mu}k_{\nu}],
\end{equation}
where we used $n^{2}=-1$. The coefficient in front is chosen such that $c_{s}=1$ leads to the standard kinetic term derived above. This operator has a two-dimensional kernel spanned by 
\begin{align}
&S_{\mu\nu}k^{\nu}=0,\qquad  \mathrm{and} \qquad S_{\mu\nu}n^{\nu}=0,
\end{align}
as $S_{\mu\nu}=0$ for $k^\mu \propto n^\mu$ ensures this second zero eigenvector.\footnote{This property explains why the speed of sound does not contribute to the advanced gauge symmetry despite it breaking Lorentz invariance. Indeed, in order to capture a unitary process, the operator must have the same kernel to the left and to the right. By breaking Lorentz invariance, one might naively expect the operator to have different left and right kernels, as it is for instance the case for the dissipative operator below. What prevents this situation to happen in the case of the speed of sound is precisely the fact the operator is annihilated by both $n^\mu$ and $k^\mu$, such that it does contribute to the noise constraint in the end.}
Note that $S_{\mu\nu}$ does not exhaust the list of all Lorentz-breaking unitary operators. For instance, at lowest order in derivatives, birefringence is also represented by a retarded gauge invariant operator linear in $k^\mu$. The new ingredient is the Levi-Civita symbol in four dimensions. In fact, the single non-zero term that involves $k^\mu$, $n^\nu$ and $\epsilon_{\mu\nu\alpha\beta}$ is gauge invariant by construction and also has a two dimensional kernel, leading to the same zero eigenvectors to the left and the right, that is
\begin{equation}
B_{\mu\nu}=i\gamma \epsilon_{\mu\nu\alpha\beta}k^{\alpha}n^{\beta}.
\end{equation}
The unitarity of the operator can be observed from the fact $B_{\mu\nu}$ is Hermitian. 

\paragraph{Lorentz violating and not unitary} The next operators we consider are those that are non-unitary and break Lorentz invariance. At lowest order in derivatives, the tensor structure is linear in $k^\mu$, such that 
\begin{equation}
    D_{\mu\nu}=\delta_{1}\omega\eta_{\mu\nu}+\delta_{2}k_{\mu}n_{\nu}+\delta_{3}n_{\mu}n_{\nu}.
\end{equation}
Again, the $\omega$ in front of $\delta_1$ comes from the contraction  $\eta_{\mu\nu}k^\mu n^\nu$. Retarded gauge invariance $D_{\mu\nu}k^{\nu}=0$ leads to
\begin{equation}
    \delta_{3}=0\;,\qquad \;\delta_{1}+\delta_{2}=0.
\end{equation}
This leaves us with a dissipation operator of the form
\begin{equation}
    D_{\mu\nu}=-i\Upgamma[\omega\eta_{\mu\nu}-k_{\mu}n_{\nu}],
\end{equation}
where we made the slight typographical change between the previously defined $\Gamma$ and this newly introduced $\Upgamma$ to stress that the former was a constant coming from the lowest order of the derivative expansion, while the latter contains higher orders in $\omega$ and $k^2$. 
This dissipative operator $D_{\mu\nu}$ is not Hermitian and therefore, the kernel to the left is not spanned by the kernel to the right and viceversa. In particular, $D_{\mu\nu}$ has $k^\mu$ as a zero eigenvector to the right and $n^\mu$ as a zero eigenvector to the left, that is
\begin{equation}
    D_{\mu\nu}k^{\nu}=0\;,\qquad \mathrm{and}\qquad \;n^{\mu}D_{\mu\nu}=0.
\end{equation}
The fact that the operator has a single zero eigenvector is essential to explain the dissipation dependence of the advanced gauge symmetry.\\


In summary, we have built the most general matrices compatible with retarded gauge invariance that includes at lowest-order in derivatives a kinetic term, non-trivial speed of sound, birefringence and dissipation, leading to
\begin{equation}
    M_{\mu\nu}=K_{\mu\nu}+S_{\mu\nu}+D_{\mu\nu}+B_{\mu\nu}.
\end{equation}
Thanks to locality, the coefficients can then be analytic functions of $\omega$ and $k^{2}$.


\paragraph{Noise constraint} We are now in the position to discuss how the kernel to the left of $M_{\mu\nu}$ leads to a constraint on the noise $\xi_{\mu}$ and the current $j_{\mu}$. To build the noise constraint, we study the kernel to the left of each of the operators that makes up the matrix $M_{\mu\nu}$, that is
\begin{align}
    &k^{\mu}K_{\mu\nu}=k^{\mu}S_{\mu\nu}=k^{\mu}B_{\mu\nu} = 0,\\
    &n^{\mu}S_{\mu\nu}=n^{\mu}B_{\mu\nu} =n^{\mu}D_{\mu\nu} = 0.
\end{align}
From the above and making use of the equations of motion $M_{\mu\nu} A^\nu= j_\mu + \xi_\mu$, we deduce that
\begin{align}
    &k^{\mu}M_{\mu\nu} A^\nu=-i\Upgamma[\omega k_{\nu}+(\omega^2 - k^2)n_{\nu}]A^{\nu}=k^{\mu}(j_{\mu}+\xi_{\mu}), \label{eq:ker1}\\
    &n^{\mu}M_{\mu\nu} A^\nu= - [\omega k_{\nu} + (\omega^2 - k^2)n_{\nu}]A^{\nu}=n^{\mu}(j_{\mu}+\xi_{\mu}). \label{eq:ker2}
\end{align}
The crucial point is that the images of $M_{\mu\nu}$ under $k^{\mu}$ and $n^{\mu}$ are proportional to each other, as it can be seen from the second equality. 
We can then find a linear combination of \Eqs{eq:ker1} and \eqref{eq:ker2} such that the two contributions cancel each other. This combination explicitly depends on the dissipation and not on any other parameters,
\begin{equation}
    (k^{\mu}-i\Upgamma n^{\mu})M_{\mu\nu}A^{\nu}=(k^{\mu}-i\Upgamma n^{\mu})(j_{\mu}+\xi_{\mu})=0\,.
\end{equation}
Therefore, in any gauge and for any $\Upgamma$, the current and stochastic noise satisfy the noise constraint
\begin{equation}\label{eq:noiseconstraintSanti}
    (k^{\mu}-i\Upgamma n^{\mu})(j_{\mu}+\xi_{\mu})=0.
\end{equation}
As expected, we recover the same result as above, from which we identify $v^\mu = k^\mu + i \Upgamma n^\mu$. 

\paragraph{Noise positivity} Before closing this section, let us comment the fact that, so far, we have not specified any property of the noise matrix $N_{\mu\nu}$ except locality. One of the non-equilibrium constraints implies that all eigenvalues of $N_{\mu\nu}$ are positive. This constraint is very general and can be made concrete by giving a general form to $N_{\mu\nu}$. We here assume that we can expand its index structure as
\begin{equation}
    N_{\alpha\beta}(k)=N_{0}(\omega,k^2)\eta_{\mu\nu}+\tilde{N}_{0}(\omega,k^2)n_{\mu}n_{\nu}+2N_{1}(\omega,k^2)k_{(\mu}n_{\nu)}+N_{2}(\omega,k^2)k_{\mu}k_{\nu}.
\end{equation}
If the matrix is positive definite, it follows from Sylvester's criterion that
\begin{align}
    &N_{0}>0,\\
    &\omega^2 N_{2} -2 \omega N_{1} +\tilde{N}_{0}>N_{0},\\
    &N_{0} N_{2} \omega^2 -(N_{0}-\tilde{N}_{0}) \left(N_{2}  k^2 +N_{0}\right)-N_{1}^2  k^2 -2 N_{0} N_{1} \omega>0.
\end{align}
A very simple way to satisfy these bounds is choosing $N_{1}=N_{2}=0$ and $\tilde{N}_{0}>N_{0}>0$.


\subsection{Covariant gauges}\label{subsec:covariant gauges}

The Coulomb gauge is not always the most convenient choice. There is a choice of gauge that retains Lorentz invariance and some gauge symmetry. These are the so-called covariant gauges. In the unitary theory they are given by introducing a new term into the action that makes the equation of motion invertible
\begin{equation}\label{eq:zetaterm}
    \mathcal{L}=-\frac{1}{4}F_{\mu\nu}F^{\mu\nu}-\frac{1}{2\zeta}\left(\partial_{\mu}A^{\mu}\right).
\end{equation}
The equations of motion for Maxwell's theory become
\begin{equation}
    \left[(-\omega^2 + k^2)\eta_{\mu\nu}-\left(1-\frac{1}{\zeta}\right)k_{\mu}k_{\nu}\right]A^{\nu}(k)=0,
\end{equation}
from which one can compute the Feynman propagator
\begin{equation}
    \langle\mathcal{T}\left\{\hat{A}^{\mu}(x)\hat{A}^{\nu}(y)\right\}\rangle=-\int\frac{d^{4}k}{(2\pi)^{4}}\frac{ie^{ik^\mu(x-y)_\mu}}{(-\omega^2 + k^2)-i\epsilon}\left[\eta^{\mu\nu}-(1-\zeta)\frac{k^{\mu}k^{\nu}}{(-\omega^2 + k^2)}\right].
\end{equation}
This propagator is not gauge invariant. The gauge invariant object is the time ordered product of the field strength

\begin{equation}
    \langle\mathcal{T}\left\{\hat{F}^{\alpha\mu}(x)\hat{F}^{\beta\nu}(y)\right\}\rangle=\langle\mathcal{T}\left\{\partial^{[\alpha}\hat{A}^{\mu]}(x)\partial^{[\beta}\hat{A}^{\nu]}(y)\right\}\rangle,
\end{equation}
where the antisymetrisation in $\alpha\mu$ and $\beta\nu$ is what removes the dependence on $\zeta$:
\begin{align}
    \langle\mathcal{T}\left\{\hat{F}^{\alpha\mu}(x)\hat{F}^{\beta\nu}(y)\right\}\rangle&=-\int\frac{d^{4}k}{(2\pi)^{4}}\frac{ie^{ik^\mu(x-y)_\mu}}{(-\omega^2 + k^2)-i\epsilon}\left[k^{[\alpha}\eta^{\mu][\nu}k^{\beta]}-(1-\zeta)\frac{k^{[\alpha}k^{\mu]}k^{[\nu}k^{\beta]}}{(-\omega^2 + k^2)}\right]\nonumber\\
    &=-\int\frac{d^{4}k}{(2\pi)^{4}}\frac{ie^{ik^\mu(x-y)_\mu}}{(-\omega^2 + k^2)-i\epsilon}\left[k^{[\alpha}\eta^{\mu][\nu}k^{\beta]}\right].\label{eq:zetaFF}
\end{align}
In this section we generalize this set of covariant gauges to Open E\&M. We start from the open effective functional constructed above. We include the $\zeta$ term of \Eq{eq:zetaterm} and compute the corresponding eigenvalues. We see that we recover two propagating degrees of freedom, a ghost and a constrained mode. Using the path integral approach, we build the retarded, advanced and Keldysh propagators for the gauge field $A^{\mu}$, both in terms of frequency and time. We prove gauge invariance of the propagators on very general grounds. Finally, we recover the noise constraint derived in \Eq{eq:noiseconstraint} in the covariant gauge formalism.


\subsubsection*{Dispersion relations}

Following the approach of \cite{Faddeev:1980be}, the variables $\zeta$ is just a number, not a field, so the inclusion into the open effective functional \eqref{OEM} is simply the term:
\begin{equation}\label{eq:zetatermra}
    -\frac{1}{2\zeta}\left(\partial_{\mu}A^{\mu}_{+}\right)^{2}+\frac{1}{2\zeta}\left(\partial_{\mu}A^{\mu}_{-}\right)^{2}=-\frac{1}{\zeta}\left(\partial_{\mu}A^{\mu}\right)\left(\partial_{\nu}a^{\nu}\right),
\end{equation}
The quadratic action is then modified to 
\begin{align}
S_1=\int \frac{\dd \omega}{2\pi} \int \frac{\dd^3 \bmk}{(2\pi)^3}   a^{\mu}M_{\mu\nu}A^{\nu}\,,
\end{align}
with
\begin{align}
M=\begin{pmatrix}  k^{2} + \frac{\omega^2}{\zeta^2}&  & - \omega k_{i} \left(1 + \frac{1}{\zeta} \right) \\ -\left( i \gamma_{2} + \frac{\omega}{\zeta} \right)k_{i} & & i\gamma_{2}\omega \delta_{ij}+\gamma_{3}(k^{2}\delta_{ij}-k_{i}k_{j})-2i\gamma_{4}\e_{ijl}k_{l}  + \frac{1}{\zeta} k_i k_j\end{pmatrix}\,.\label{eq:zetaMmatrix}
\end{align}
The associated eigenvalues are
\begin{align}\label{eq:eigenMnew} 
\{C_\zeta( 1 + \delta_\zeta) , C_\zeta (1- \delta_\zeta), 
k^2 \gamma_3 - 2 k \gamma_4 + i \gamma_2 \omega, 
k^2 \gamma_3 + 2 k \gamma_4 + i \gamma_2 \omega \}\,,
\end{align}
with
\begin{align}
    C_\zeta &\equiv \frac{1}{2}\left(k^2 + i \gamma_2 \omega\right) + \frac{1}{2\zeta}\left( k^2 + \omega^2\right) \,,\\
    \delta_\zeta &\equiv  \sqrt{1 - 4 \frac{\zeta \left(k^2 - \omega^2 \right) \left( k^2 - i \gamma_2 \omega\right)}{\left[ \left(k^2 + \omega^2 \right) + \zeta\left(k^2 + i \gamma_2 \omega\right)\right]^2}}.
\end{align}
This result has to be compared to our previous finding in \Eq{eq:eigenM}. First, we notice that that the determinant of $M$ does not vanish anymore, 
\begin{align}
    \det M =  \frac{1}{\zeta} (k^2 - \omega^2 )  \left( k^2 - i \gamma_2 \omega \right) \left(i \gamma_2 \omega -2 \gamma_4 k +\gamma_3 k^2\right) \left(i \gamma_2 \omega + 2 \gamma_4 k +\gamma_3 k^2\right).
\end{align}
Hence, for $\zeta \neq 0$, $M$ might be directly inverted to obtain the propagators in the covariant gauge. Second, comparing \Eq{eq:eigenM} with \Eq{eq:eigenMnew}, the two propagating modes related to the eigenvalues $\lambda_\pm = k^2 \gamma_3 \pm 2 k \gamma_4 + i \gamma_2 \omega$ are unchanged. Third, we deduce that the last two eigenvalues $ C_\zeta (1 \pm \delta_\zeta)$ must then be associated with the ghost and the constrained mode. 

\subsubsection*{Propagators}

The path integral approach is extremely convenient to derive an expression for the Keldysh propagator without having to use the Langevin equation. To compute the Keldysh propagator, we introduce two current that couple to the advanced and retarded components of the field. We compute the Gaussian path integral and find a partition function. This yields the answer for the Keldysh and retarded propagator in Fourier space. Then we can perform the frequency integral to find the time-momentum representation. Consider the partition function
\begin{equation}
    \mathcal{Z}[J^{A},J^{a}]=\int\mathcal{D}A^{\mu}(t_{f})\int_{\text{BD}}^{A^{\mu}(t_{f})}\mathcal{D}A^{\mu}(t)\int_{\text{BD}}^{0}\mathcal{D}a^{\mu}(t)\,e^{iS[A,a]+\int J^{a}_{\mu}a^{\mu}+\int J^{A}_{\mu}A^{\mu}}
\end{equation}
where the open effective functional is
\begin{equation}
    iS[A,a]=\frac{1}{2}\int\frac{d^{4}k}{(2\pi)^{4}}\left(A^{\mu}(-\omega, - \bmk),a^{\mu}(-\omega, - \bmk)\right)\begin{pmatrix}
        0 & i M_{\nu\mu}(-\omega, -\bmk)\\
        i M_{\mu\nu}(\omega, \bmk) & -2N_{\mu\nu}(\omega, \bmk)
    \end{pmatrix}\begin{pmatrix}
        A^{\nu}(\omega, \bmk)\\
        a^{\nu}(\omega, \bmk)
    \end{pmatrix},
\end{equation}
and the coupling to the sources
\begin{align}
    \int d^{4}x J^{a}_{\mu}(x)a^{\mu}(x)=&\frac{1}{2}\int\frac{d^{4}k}{(2\pi)^{4}}\left[J^{a}_{\mu}(-\omega, -\bmk)a^{\mu}(\omega, \bmk)+a^{\mu}(-\omega, -\bmk)J^{a}_{\mu}(\omega, \bmk)\right],\\
    \int d^{4}x J^{A}_{\mu}(x)A^{\mu}(x)=&\frac{1}{2}\int\frac{d^{4}k}{(2\pi)^{4}}\left[J^{A}_{\mu}(-\omega, -\bmk)A^{\mu}(\omega, \bmk)+A^{\mu}(-\omega, -\bmk)J^{A}_{\mu}(\omega, \bmk)\right].
\end{align}
Note that $N_{\mu\nu}(\omega, \bmk)=N_{\nu\mu}(\omega, \bmk)=N_{\mu\nu}(-\omega,-\bmk)$ by construction\footnote{To be precise, only the symmetric and even part of the noise matrix enters the equation. Its symmetric part is picked up by the fact that we are taking $aNa$ with $a^{\mu}(x)$ evaluated at the same point. The even part is coming from the fact that we can split $2aNa=a(-k)N(k)a(k)+a(k)N(k)a(-k)$ and then change the variables of integration in the latter and use the fact that $N$ is symmetric to write $2aNa=a(-k)N(k)a(k)+a(-k)N(-k)a(k)$.}. To perform the path integral, it is necessary to do a shift of the retarded and advanced components
\begin{align}
    &A^{\mu}(x)\rightarrow A^{\mu}(x)+\int d^{4}z\left[R^{\mu\alpha}_{Aa}(x-z)J^{\alpha}_{a}(z)+R^{\mu\alpha}_{AA}(x-z)J^{\alpha}_{A}(z)\right],\\
    &a^{\mu}(x)\rightarrow a^{\mu}(x)+\int d^{4}z\left[R^{\mu\alpha}_{aA}(x-z)J^{\alpha}_{A}(z)\right].
\end{align}
The preservation of the boundary conditions of the advanced component means that $R^{\mu\alpha}_{aA}$ has advanced boundary conditions. We use this shift to remove any of the terms linear in the sources in the integral. This leads to a set of three equations in Fourier space
\begin{align}
    &iM_{\alpha\mu}(-\omega, -\bmk)R^{\alpha\nu}_{aA}(\omega,\bmk)+\delta^{\nu}_{\mu}=0,\\
    &iM_{\alpha\mu}(\omega,\bmk)R^{\alpha\nu}_{Aa}(\omega,\bmk)+\delta^{\nu}_{\mu}=0,\\
    &iM_{\mu\alpha}(\omega, \bmk)R^{\alpha\nu}_{AA}(\omega, \bmk)-2N_{\mu\alpha}(\omega, \bmk)R^{\alpha\nu}_{aA}(\omega, \bmk)=0.
\end{align}
which solve to
\begin{align}
    &R^{\alpha\nu}_{aA}(\omega, \bmk)=i[M(-\omega, -\bmk)^{-1}]^{\nu\alpha},\\
    &R^{\alpha\nu}_{Aa}(\omega, \bmk)=i[M(\omega, \bmk)^{-1}]^{\alpha\nu},\\
    &R^{\alpha\nu}_{AA}(\omega, \bmk)=2i[M(\omega, \bmk)^{-1}]^{\alpha\rho}[M(-\omega, -\bmk)^{-1}]^{\nu\sigma}N_{\rho\sigma}(\omega, \bmk).
\end{align}
The final result for the partition function is:
\begin{equation}
    Z[J^{A},J^{a}]=Z[0,0]\text{exp}\left\{\frac{1}{2}\int\frac{d^{4}k}{(2\pi)^{4}}\left(J^{A}_{\mu}(-\omega, -\bmk)J^{a}_{\mu}(-\omega, -\bmk)\right)H^{\mu\nu}(\omega, \bmk)\begin{pmatrix}
        J^{A}_{\nu}(\omega, \bmk)\\
        J^{a}_{\nu}(\omega, \bmk)
    \end{pmatrix}\right\},
\end{equation}
where
\begin{equation}\label{eq:matrixH}
     H^{\mu\nu}=\begin{pmatrix}
      2[M(\omega, \bmk)^{-1}]^{\mu\alpha}[M(-\omega, -\bmk)^{-1}]^{\nu\beta}\left[N_{\alpha\beta}(\omega, \bmk)\right] & i[M(\omega, \bmk)^{-1}]^{\mu\nu}\\
      i[M(-\omega, -\bmk)^{-1}]^{\nu\mu}  & 0
    \end{pmatrix}.
\end{equation}
If we now provide an explicit expression for $M^{-1}$ and $N$, we can extract the expression of the propagators in the covariant gauge. 


\paragraph{Retarded propagator.} Inverting the matrix in full generality may be cumbersome, especially when all EFT coefficients are non-vanishing. Yet, analytic results can be reached, following the constructive approach of \Sec{subsec:SantiEM}. 
The retarded propagator\footnote{Since the retarded propagator is independent of state, one can remove the expectation value on the operator expression at the price of introducing an identity operator \cite{breuerTheoryOpenQuantum2002}.}
\begin{align}
    G_{R}^{\mu\nu}(x-y)&=\langle[\hat{A}^{\mu}(x),\hat{A}^{\nu}(y)]\rangle\theta(x^{0}-y^{0}) = \exx{A^{\mu}(x) a^{\nu}(y)}
\end{align}
is given by the Fourier transform of $R_{Aa}^{\mu\alpha}$, that is
\begin{align}
    G_{R}^{\mu\nu}(x-y) &=2\int\frac{d^{4}k}{(2\pi)^{4}}i[M(-\omega, -\bmk)^{-1}]^{\mu\nu}e^{ik^\alpha(x-y)_\alpha}, \\
    &\equiv \int\frac{d^{4}k}{(2\pi)^{4}}G_{R}^{\mu\nu}(\omega, \bmk)e^{ik^\alpha(x-y)_\alpha}.
\end{align}
We need to find the inverse matrix of $M(-\omega, -\bmk)$. Following \Sec{subsec:SantiEM}, the matrix takes the explicit form 
\begin{align}
    M_{\mu\nu}(\omega, \bmk)&=(-\omega^2 + k^2)\eta_{\mu\nu}-\left(1-\frac{1}{\zeta}\right)k_{\mu}k_{\nu}-i\Upgamma[\omega \eta_{\mu\nu}-k_{\mu}n_{\nu}]+i\gamma_{4}\epsilon_{\mu\nu\alpha\beta}k^{\alpha}n^{\beta} \nonumber \\
    +&(c_{s}^{2}-1)[k^{2}n_{\mu}n_{\nu}-\omega(k_{\mu}n_{\nu}+n_{\mu}k_{\nu})+ k^2\eta_{\mu\nu}-k_{\mu}k_{\nu}],
\end{align}
where we considered dispersion ($c_s \neq 1$), dissipation ($\gamma \neq 0$) and birefringence ($\gamma_4 \neq 0$) of the medium. The inverse is also a linear combination of similar terms since they span the range of allowable matrices. We define
\begin{equation}\label{eq:retfull}
   G_{R}^{\mu\nu}(\omega, \bmk)=2i \left(A\eta^{\mu\nu}+Bk^{\mu}k^{\nu}+Ck^{\mu}n^{\nu}+Dn^{\nu}k^{\mu}+E\epsilon^{\mu\nu\alpha\beta}k_{\alpha}n_{\beta}+G n^{\mu}n^{\nu}\right).
\end{equation}
for some unknown coefficients $A$, $B$, $C$, $D$, $E$ and $G$. Solving the equation $M^{-1}M=\mathbb{I}$ is complicated for general $\zeta$. However, the derivative of $M^{-1}$ in terms of zeta is simple, 
\begin{equation}
    \partial_{\zeta}[M^{-1}]^{\mu\nu}=-M^{1}\partial_{\zeta}M M^{-1}=\frac{k^{\mu}k^{\nu}+i\Upgamma\omega k^{\mu}n^{\nu}}{(-\omega^2 + k^2)[(-\omega^2 + k^2)+i\Upgamma \omega]}.
\end{equation}
This allows us to invert $M$ by matching at $\zeta=0$. To compute the general expression for the coefficients in \eqref{eq:retfull} as functions of $\omega$ and $\bfk$ we can make use the rotational invariance of the theory. We perform the matching in the reference frame $k^{\mu}=(\omega,0,0,k)$ and then extend the dependence to an arbitrary frame $k^{\mu}=(\omega,\bfk)$. One finds the generic coefficients to be:
\begin{align}
    A&=- \frac{-\omega^2 +i\Upgamma\omega +  c_{s}^{2} k^2}{(-\omega^2 +i\Upgamma\omega +  c_{s}^{2} k^2)^{2} -\gamma_{4}^{2}k^2}, \Bigg.\\
    B&=-\frac{\omega^{2}(1 + \zeta) + \zeta k^2}{(-\omega^2 + k^2)(k^2+\omega^{2})(-\omega^2 + i\Upgamma \omega + k^2)}-\frac{-\omega^2 +i\Upgamma\omega +  c_{s}^{2} k^2}{k^2[(-\omega^2 +i\Upgamma\omega +  c_{s}^{2} k^2)^{2} -\gamma_{4}^{2}k^2]}, \Bigg.\\
    C&=\frac{\omega(-\omega^2 +i\Upgamma\omega +  c_{s}^{2} k^2)}{k^2[(-\omega^2 +i\Upgamma\omega +  c_{s}^{2} k^2)^{2} -\gamma_{4}^{2}k^2]} + \frac{\omega[-\omega^2 + (1 + i\Upgamma\zeta) k^2]}{k^2(-\omega^2 + k^2)(-\omega^2 +i\Upgamma\omega+ k^2)}, \Bigg.\\
    D&=\frac{\omega(-\omega^2 +i\Upgamma\omega +  c_{s}^{2} k^2)}{k^2[(-\omega^2 +i\Upgamma\omega +  c_{s}^{2} k^2)^{2} -\gamma_{4}^{2}k^2]}  +\frac{\omega}{(-\omega^2 +i\Upgamma \omega+ k^2)(k^2+\omega^{2})}, \Bigg.\\
    E&=\frac{i\gamma_{4}}{(-\omega^2 +i\Upgamma\omega +  c_{s}^{2} k^2)^{2} -\gamma_{4}^{2}k^2}, \Bigg. \\
    G&=\frac{(-\omega^2 + k^2)(-\omega^2 +i\Upgamma\omega +  c_{s}^{2} k^2)}{k^2[(-\omega^2 +i\Upgamma\omega +  c_{s}^{2} k^2)^{2} -\gamma_{4}^{2}k^2]} +\frac{(-\omega^2 + k^2)}{k^2(-\omega^2 +i \Upgamma \omega+ k^2)} \Bigg. .
\end{align}
Interestingly, among all these coefficients, only $B$ and $C$ feature $\zeta$. \Eq{eq:retfull} provides a complete characterization of the retarded Green's function in the covariant gauges, in the presence of dissipation, speed of sound and birefringence.

\paragraph{Keldysh propagator.} The Keldysh propagator depends on the noise kernel. Using that $N$ is symmetric leads to the parametrization
\begin{equation}
    N_{\alpha\beta}(\omega, \bmk)+N_{\beta\alpha}(-\omega, -\bmk)=2N_{0}(\omega^{2},k^{2})\eta_{\mu\nu}+2\tilde{N}_{0}(\omega^{2},k^{2})n_{\mu}n_{\nu}+2N_{2}(\omega^{2},k^{2})k_{\mu}k_{\nu}.
\end{equation}
Hence, the Keldysh propagator 
\begin{align}
    G_{K}^{\mu\nu}(x-y)=&\frac{1}{2}\langle\{\hat{A}^{\mu}(x),\hat{A}^{\nu}(y)\}\rangle= \exx{A^{\mu}(x) A^\nu(y)}  
\end{align}
expressed in Fourier space is
\begin{align}
    G_{K}^{\mu\nu}(x-y)&=\int\frac{d^{4}k}{(2\pi)^{4}}ie^{ik^\alpha(x-y)_\alpha}\left\{[M(\omega, \bmk)^{-1}]^{\mu\alpha}[M(-\omega, -\bmk)^{-1}]^{\nu\beta}\left[N_{\alpha\beta}(\omega, \bmk)+N_{\beta\alpha}(-\omega, -\bmk)\right]\right\}, \nonumber \\
    &\equiv \int\frac{d^{4}k}{(2\pi)^{4}}G_{K}^{\mu\nu}(\omega, \bmk)e^{ik^\alpha(x-y)_\alpha}.
\end{align}
The Keldysh propagator in Fourier space can only comprise of symmetric tensors:
\begin{equation}
     G_{K}^{\mu\nu}(\omega, \bmk)=i\sum_{n}K_{n}^{\mu\nu}(\omega, \bmk),
\end{equation}
which must verify the ``symmetric-even" property
\begin{equation}
    \text{Symmetric-even: } \quad K_{n}^{\mu\nu}(\omega, \bmk)=K_{n}^{\nu\mu}(-\omega, -\bmk).
\end{equation}
This is a consequence of the expression we found for the Keldysh propagator in \eqref{eq:matrixH}. We can decompose the Keldysh propagator into six symmetric-even tensors as
\begin{equation}\label{eq:keldyshinv}
    G_{K}^{\mu\nu}(\omega, \bmk)= i \left[\tilde{A}\eta^{\mu\nu}+\tilde{B}k^{\mu}k^{\nu}+\tilde{C}_{1}\left(k^{\mu}n^{\nu}+k^{\nu}n^{\mu}\right)+\tilde{C}_{2}\left(k^{\mu}n^{\nu}-k^{\nu}n^{\mu}\right)+\tilde{D}n^{\mu}n^{\nu}+\tilde{E}\epsilon^{\mu\nu\alpha\beta}k_{\alpha}n_{\beta}\right], \Bigg.
\end{equation}
where $\tilde{A},\tilde{B},\tilde{C}_{2},\tilde{D}$ and $\tilde{E}$ are even functions of $\omega$ and $\bfk$, while $\tilde{C}_{1}$ is an odd function of $\omega$ and $\bfk$. Thanks to the spatial rotation invariance of the theory we can choose the reference frame where $k^{\mu}=(\omega,0,0,k)$ and match the left-hand side and the right-hand side of \Eq{eq:keldyshinv}. The expressions we derive for the coefficient in \Eq{eq:keldyshinv} can then be extended to a general reference frame $k^{\mu}=(\omega,\bfk)$, leading to
\begin{align}
    \tilde{A}=&\frac{2 N_{0} \left[\left(\omega^2-c_{s}^2 k^2\right)^2+\Upgamma^2 \omega^2+\gamma_{4}^2 k^2\right]}{\left[\left(- \omega^2 +i  \Upgamma \omega + c_{s}^2 k^2\right)^2 - \gamma_{4}^2 k^2\right] \left[\left(-\omega^2 -i \Upgamma \omega +c_{s}^2 k^2\right)^2 - \gamma_{4}^2 k^2\right]}, \Bigg.
    \\
    \tilde{B}=
    &-\frac{2 N_{0} \left[\Upgamma^2 \omega^2+ \left(\omega^2-c_{s}^2 k^2\right)^2+\gamma_{4}^2 k^2\right]}{k^2\left[\left(- \omega^2 +i  \Upgamma \omega + c_{s}^2 k^2\right)^2 - \gamma_{4}^2 k^2\right] \left[\left(-\omega^2 -i \Upgamma \omega +c_{s}^2 k^2\right)^2 - \gamma_{4}^2 k^2\right]}   \Bigg. \nonumber
     \\
    &+ \frac{2 N_{0} \omega^2\left(\omega^2 - k^2\right) + 2 \tilde{N}_{0}  \omega^2  k^2  }{k^2 \left(\omega^2 - k^2\right)^2 \left[ \Upgamma^2 \omega^2 +  \left(\omega^2 -k^2\right)^2\right]} -  \frac{ 4 \tilde{N}_{0}  \omega^2  k^2 \zeta  }{k^2 \left(\omega^2 - k^2\right)^2 \left[ \Upgamma^2 \omega^2 +  \left(\omega^2 -k^2\right)^2\right]} \Bigg. \nonumber
     \\
    &+\frac{2 \zeta^2 \left\{N_{0} \left(-\omega^2 - \Upgamma^2 + k^2 \right) + \tilde{N}_0\left( \omega^2 + \Upgamma^2\right) + N_2 \left[\Upgamma^2\omega^2 + \left(k^{2}-\omega^2\right)^2 \right]\right\}}{\left(k^{2}-\omega^2\right)^2 \left[\Upgamma^2\omega^2 + \left(k^{2}-\omega^2\right)^2 \right]}, \Bigg. 
    \\
    \tilde{C}_{1}=&\frac{ 2 \omega \left[ N_{0} \left(-\omega^2 + k^2 \right) - \tilde{N}_{0} k^2 \right]}{k^2 \left(\omega^2 - k^2\right)^2 \left[ \Upgamma^2 \omega^2 +  \left(\omega^2 -k^2\right)^2\right]} 
    +
    \frac{ 2 \tilde{N}_{0} \omega \zeta}{ \left(\omega^2 - k^2\right)^2 \left[ \Upgamma^2 \omega^2 +  \left(\omega^2 -k^2\right)^2\right]}\Bigg. \nonumber
    \\
    &-\frac{2 N_{0} \omega \left[\Upgamma^2 \omega^2+ \left(\omega^2-c_{s}^2 k^2\right)^2+\gamma_{4}^2 k^2\right]}{k^2 \left[\left(- \omega^2 +i  \Upgamma \omega + c_{s}^2 k^2\right)^2 - \gamma_{4}^2 k^2\right] \left[\left(-\omega^2 -i \Upgamma \omega +c_{s}^2 k^2\right)^2 - \gamma_{4}^2 k^2\right]},\Bigg. 
    \\
    \tilde{C}_{2}=&\frac{2 i \Upgamma \zeta (N_{0}-\tilde{N}_{0})}{(k^{2}-\omega^{2}) \left[ \Upgamma^2 \omega^2 +  \left(\omega^2 -k^2\right)^2\right]}, \Bigg.
    \\
    \tilde{D}=&-\frac{2 N_{0} (\omega^{2} - k^2) \left[\Upgamma^2 \omega^2+ \left(\omega^2-c_{s}^2 k^2\right)^2+\gamma_{4}^2 k^2\right]}{k^2 \left[\left(- \omega^2 +i  \Upgamma \omega + c_{s}^2 k^2\right)^2 - \gamma_{4}^2 k^2\right] \left[\left(-\omega^2 -i \Upgamma \omega +c_{s}^2 k^2\right)^2 - \gamma_{4}^2 k^2\right]}\Bigg. \nonumber \\
    &+\frac{2 N_{0} \left(\omega^2 - k^2\right) + 2 k^2 \tilde{N}_{0}}{k^2 \left[ \Upgamma^2 \omega^2 +  \left(\omega^2 -k^2\right)^2\right]}, \Bigg.
    \\
    \tilde{E}=&-\frac{4 i \gamma_{4} N_{0} \left(\omega^{2} - c_{s}^{2} k^{2}\right) }{\left[\left(- \omega^2 +i  \Upgamma \omega + c_{s}^2 k^2\right)^2 - \gamma_{4}^2 k^2\right] \left[\left(-\omega^2 -i \Upgamma \omega +c_{s}^2 k^2\right)^2 - \gamma_{4}^2 k^2\right]}. \Bigg.
\end{align}
Again, only $\tilde{B}$ and $\tilde{C}_{1,2}$ depend on $\zeta$. Furthermore, the only odd function of $\omega$ and $\bfk$ is $\tilde{C}_{1}$, as expected. Note that computing gauge invariant quantities requires multiplying by $k^{\alpha}k^{\beta}$ and antisymmetrising the pairs $[\alpha\mu]$ and $[\beta\nu]$, since one must consider the Keldysh propagator of the retarded field strength. This removes any contribution from $\tilde{B}$ or $\tilde{C}_{1,2}$, which were also the only ones to include any $\zeta$ dependence. This is a direct way to see that indeed the out-of-time ordered correlation function of the field strength of the retarded gauge field is $\zeta$ independent as it should be. Furthermore, this proves that $N_{2}$ does not contribute to the field strength, hence it would be enough to only consider the pair $N_{0}$ and $\tilde{N}_{0}$. One can then greatly reduce the structure of the noise matrix $N$ without losing any physics. The positivity condition then reduce to
\begin{equation}
    N_{0}>0\;,\;\quad\tilde{N}_{0}-N_{0}>0,
\end{equation}
which is much more compact than before.


\subsubsection*{Gauge invariance}

We can now show that the retarded, advanced and Keldysh propagators of the field strength are gauge invariant as they should be. Using the notation of Section \Sec{subsec:path}, gauge invariance implies that
\begin{align}
    \partial_{\zeta}\langle[\hat{F}^{\mu\nu}(t_{1},\bmk_{1}),\hat{A}^{\rho}(t_{2},\bmk_{2})]\rangle\theta(t_{1}-t_{2})&= \partial_{\zeta}\exx{F^{\mu\nu}(t_{1},\bmk_{1}) a^{\rho}(t_{2},\bmk_{2})}=0,\\
    \partial_{\zeta}\langle \{\hat{F}^{\mu\nu}(t_{1},\bmk_{1}),\hat{F}^{\mu\nu}(t_{2},\bmk_{2})\}\rangle &= 2\partial_{\zeta}\exx{F^{\mu\nu}(t_{1},\bmk_{1})F^{\rho\sigma}(t_{2},\bmk_{2}}=0.
\end{align}
These relations are easier to prove in Fourier space. We start from the retarded propagator obtained from
\begin{align}
    G_{R}^{\mu\nu}(x-y)&=\langle[\hat{A}^{\mu}(x), \hat{A}^{\nu}(y)]\rangle\theta(x^{0}-y^{0}) = \exx{A^{\mu}(x) a^\nu(y)} \\
    &=2\int\frac{d^{4}k}{(2\pi)^{4}}i[M(-\omega, -\bmk)^{-1}]^{\mu\nu}e^{ik^\alpha(x-y)_\alpha}.
\end{align}
The gauge invariant object is given by a derivative acting on the retarded component
\begin{align}
    \langle[\hat{F}^{\alpha\mu}(x),\hat{A}^{\nu}(y)]\rangle\theta(x^{0}-y^{0}) &= \exx{F^{\alpha\mu}(t_{1},\bmk_{1}) a^{\nu}(t_{2},\bmk_{2})} \\
    &=-2\int\frac{d^{4}k}{(2\pi)^{4}}k^{[\alpha}[M(-\omega, -\bmk)^{-1}]^{\mu]\nu}e^{ik^\beta(x-y)_\beta},
\end{align}
such that a sufficient condition for gauge invariance is
\begin{equation}
    \partial_{\zeta}\left(k^{[\alpha}[M(-\omega, -\bmk)^{-1}]^{\mu]\nu}\right)=0.
\end{equation}
We can use the chain rule to rewrite the derivative on $M^{-1}$ in terms of the derivative acting on $M$ to obtain
\begin{equation}
    k^{[\alpha}\partial_{\zeta}[M(-\omega, -\bmk)^{-1}]^{\mu]\nu}=-k^{[\alpha}[M(-\omega, -\bmk)^{-1}]^{\mu]\rho}\partial_{\zeta}M_{\rho\sigma}(-\omega, -\bmk)[M(-\omega, -\bmk)^{-1}]^{\sigma\nu}.
\end{equation}
Then, the derivative with respect to $\zeta$ is given by
\begin{equation}\label{eq:simplif}
    \partial_{\zeta}M_{\rho\sigma}(-\omega, -\bmk)=\frac{1}{\zeta^{2}}k_{\rho}k_{\sigma},
\end{equation}
such that the action of the inverse matrix on $k$ is equivalent to solving the equation for a vector $b$ obeying $M_{\mu\nu}b^{\nu}=k_{\mu}$. The solution is
\begin{equation}
    b^{\nu}=\frac{\zeta}{k^{2}}k^{\nu},
\end{equation}
from which the sufficient condition for gauge invariance simplifies to
\begin{equation}
   -\frac{1}{\zeta k^{2}}k^{[\alpha}k^{\mu]}k_{\sigma}[M(-\omega, -\bmk)^{-1}]^{\sigma\nu}=0.
\end{equation}
This expression manifestly vanishes from the antisymmetrisation of $k^{[\alpha}k^{\mu]}$. 

This proof of gauge invariance can also be applied to the Keldysh propagator
\begin{align}
    &\frac{1}{2}\langle\{\hat{A}^{\mu}(x),\hat{A}^{\nu}(y)\}\rangle= \exx{A^{\mu}(x) A^\nu(y)}  \\
    =&\int\frac{d^{4}k}{(2\pi)^{4}}ie^{ik^\rho(x-y)_\rho}\left\{[M(\omega, \bmk)^{-1}]^{\mu\alpha}[M(-\omega, -\bmk)^{-1}]^{\nu\beta}\left[N_{\alpha\beta}(\omega, \bmk)+N_{\beta\alpha}(-\omega, -\bmk)\right]\right\}.
\end{align}
There, a sufficient condition for gauge invariance is given by
\begin{equation}
    \partial_{\zeta}\left(k^{[\alpha}[M(\omega, \bmk)^{-1}]^{\mu]\rho}k^{[\beta}[M(-\omega, -\bmk)^{-1}]^{\nu]\sigma}\left[N_{\rho\sigma}(\omega, \bmk)+N_{\sigma\rho}(-\omega, -\bmk)\right]\right)=0.
\end{equation}
It turns out this equation is automatically verified thanks to the gauge invariance of the retarded propagator and the advanced propagator proven before. 


\subsubsection*{Noise constraint}

One of the main results from Sec. \ref{subsec:noise} is the constraint associated to the advanced gauge transformation, which we dubbed \textit{noise constraint},
\begin{equation}
    v^{\mu}(\xi_{\mu}+j_{\mu})=0.
\end{equation}
This constraint arises from the existence of a zero eigenvector for the matrix \eqref{eq:matrix1}. However, upon the introduction of the $\zeta$ term \eqref{eq:zetatermra}, this zero eigenvector disappears. Therefore, the noise constraint has to be recovered in a different way when working in covariant gauges. It turns out that one can make it explicit by using a quantisation condition similar to Gupta-Bleuler condition \cite{Gupta:1949rh, Bleuler:1950cy}
\begin{equation}\label{eq:GBcondition}
   \bra{\Psi} \partial^{\mu}\hat{A}_{\mu}\ket{\Psi}=0,
\end{equation}
where $\ket{\Psi}$ is a physical state. The analogous condition for open E$\&$M is given in terms of the expectation value 
for the retarded field
\begin{equation}\label{eq:newGPcondition}
    \;\forall n\in\mathbb{N}\;,\;\exx{\left(\partial^{\mu}A_{\mu}\right)^{n}}=0,
\end{equation}
where the double brackets stands for an average over the noise, as defined in \Sec{subsec:path}. 

Let us derive the noise constraint from this condition. First, we can solve the equations of motion in Fourier space by inverting the matrix $M$, which yields
\begin{equation}
    A^{\mu}(\omega, \bmk)=\left[M^{-1}(\omega, \bmk)\right]^{\mu\nu}\xi_{\nu}(\omega, \bmk).
\end{equation}
Taking the $n=1$ case of the new condition \eqref{eq:newGPcondition}, we find
\begin{equation}
    \left\langle\left\langle\partial^{\mu}A_{\mu}\right\rangle\right\rangle=-i\left\langle\left\langle k_{\mu}\left[M^{-1}(\omega, \bmk)\right]^{\mu\nu}\xi_{\nu}(\omega, \bmk)\right\rangle\right\rangle=0.
\end{equation}
As above, we need to compute $k_{\mu}\left[M^{-1}(\omega, \bmk)\right]^{\mu\nu}$, which is equivalent to find $b^\nu$ such that $k_{\mu}=b^{\nu}M_{\nu\mu}$.
To achieve this inversion, we decompose $b^{\nu}=A k^{\nu}+B n^{\nu}$ such that
\begin{align}
    b^{\nu}M_{\nu\mu}&= A \left\{ \frac{1}{\zeta}(-\omega^2 + k^2) k_{\mu}+(-i\Upgamma)\left[\omega k_{\nu}-(-\omega^2 + k^2)n_{\nu}\right]\right\}\nonumber \\
    &\qquad +B\left[(-\omega^2 + k^2)n_{\nu}-(1-\frac{1}{\zeta})\omega k_{\nu}\right].
\end{align}
The key part is finding the cancellation of the $n_{\nu}$ terms, which leaves
\begin{equation}
    i\Upgamma A (-\omega^2 + k^2) n_{\nu}+B (-\omega^2 + k^2) n_{\nu}=0 \quad \rightarrow \quad  B=-iA\Upgamma.
\end{equation}
Consequently, the expression simplifies to 
\begin{equation}
    b^{\nu}M_{\nu\mu}=A \left[\frac{1}{\zeta} (-\omega^2 + k^2) k_{\nu}-\frac{i\Upgamma}{\zeta} \omega k_{\nu} \right] = k_{\nu}\quad \rightarrow \quad b^{\nu}=\frac{\zeta(k^{\nu}-i\Upgamma n^{\nu})}{ -\omega^2 -i\Upgamma \omega + k^2}.
\end{equation}
Crucially, this implies that $b^{\nu}$ is proportional to the generator of the advanced gauge transformation found in \Sec{subsec:SantiEM}. For the $n=1$ case of the condition givein in \Eq{eq:newGPcondition}, we then have
\begin{equation}
    \left\langle\left\langle\partial^{\mu}A_{\mu}\right\rangle\right\rangle=-\frac{i\zeta}{-\omega^2 -i\Upgamma \omega + k^2}\left\langle\left\langle (k^{\nu}-i\Upgamma n^{\nu})\xi_{\nu}\right\rangle\right\rangle=0,
\end{equation}
which exactly recovers the noise constraint we found the constructive approach of \Eq{eq:noiseconstraintSanti} through the identification $v^\nu = k^{\nu}-i\Upgamma n^{\nu}$. One can further take the $n=2$ case of the constraint to find
\begin{equation}
    \exx{\left(\partial^{\mu}A_{\mu}\right)^{2}}=0\quad \rightarrow \quad v^{\mu}v^{\nu}\exx{\xi_{\mu}\xi_{\nu}}=0,
\end{equation}
which is nothing but the condition derived in \Eq{eq:noisePk} for the power spectrum of the noise entering the Langevin equation.


\subsection{Topological operators}\label{subsec:topological}

Allowing for dissipation and noise enlarge the class of operators accessible from within the EFT. Do some of these new operators have particular properties? In the unitary theory, a well known non-trivial extension of Maxwell theory consists in adding a topological operator known as \textit{theta term}
\begin{align}
    S_{\theta_1} = \frac{\theta_1}{4} \int \dd^4x ^\star\!F^{\mu\nu} F_{\mu\nu} = \frac{\theta_1}{4} \int \dd^4x \bm{\mathrm{E}}.\bm{\mathrm{B}}
\end{align}
where 
\begin{align}
     ^\star\! F^{\mu\nu} = \frac{1}{2} \epsilon^{\mu\nu\rho\sigma}F_{\rho\sigma}.
\end{align}
When $\theta_1$ is a constant, it is easy to check $S_{\theta_1}$ turns out to be a total derivative, 
\begin{align}
    S_{\theta_1} = \frac{\theta_1}{8} \int \dd^4x \partial_\mu\left(\epsilon^{\mu\nu\rho\sigma}A_\nu\partial_{\rho}A_\sigma\right),\label{eq:thetauni}
\end{align}
such that all the physical information is encoded in the spacetime boundary. Are there any non-unitary operators that exhibit similar properties? 

\subsubsection*{Linear in the advanced field}

It is first instructive to express $S_{\theta_1}$ in the Keldysh basis,
\begin{align}\label{eq:top1}
    S_{\theta_1}^+ - S_{\theta_1}^- = \frac{\theta_1}{4} \int \dd^4x  ^\star\! F^{\mu\nu} \partial_\mu a_\nu,
\end{align}
where we used the same notations as above for the retarded field strength $F^{\mu\nu}$ and the advanced component $a_\nu$. Using Bianchi identity
\begin{align}
     \partial_\mu \, ^\star\! F^{\mu\nu} = 0,
\end{align}
it is then straightforward to check this term is a total derivative. At linear order in the advanced field, \Eq{eq:top1} is the only topological operator one can construct. Indeed, its topological nature deeply relates  to the use of the Levi-Civita symbol $\epsilon^{\mu\nu\rho\sigma}$. The operator being linear in $A_\rho$ and $a_\sigma$, the only non-vanishing way to contract the remaining indices consists in acting with one derivative on each field, and so the result of \Eq{eq:top1}. Can we then construct new topological operators in the open theory?

\subsubsection*{Quadratic in the advanced field}

A direct way to construct a new topological operator consists in considering the analogue of the theta term in the advanced sector. By introducing 
\begin{align}
    F_{a,\mu\nu} = \partial_\mu a_\nu - \partial_\nu a_\mu, \quad \mathrm{and} \quad ^\star\! F^{\mu\nu}_a = \frac{1}{2} \epsilon^{\mu\nu\rho\sigma}F_{a,\rho\sigma},
\end{align}
we construct
\begin{align}
    S_{\theta_2} = i \frac{\theta_2}{4} \int \dd^4x  ^\star\! F^{\mu\nu}_a \partial_\mu a_\nu.
\end{align}
This term is a topological noise operator, which is indeed a total derivative
\begin{align}
    S_{\theta_2} = i\frac{\theta_2}{8} \int \dd^4x \partial_\mu\left(\epsilon^{\mu\nu\rho\sigma}a_\nu\partial_{\rho}a_\sigma\right).\label{eq:thethanoise}
\end{align}
Note that convergence of the path integral necessitates $\Ima S_{\mathrm{eff}}\geq 0$ for all field values, which might be a non-trivial constraint. One also must bear in mind that the advanced field vanishes at the turn-around time $t_f$ of the closed-time path contour, $a_\nu(t_f,\bmx) = 0$. This boundary condition might restrict the impact of such operator on the diagonal element of the density matrix probed by the path integral (but may nevertheless affect off-diagonal density matrix elements obeying different boundary conditions). At last, if $\theta_2$ varies in space and time, that is $\theta_2 = \theta_2(t, \bmx)$, the integration by part leads to novel contributions to stochastic part of the equations of motion. We leave their investigation and eventual connections with chiral vacuum memory effect \cite{Maleknejad:2023nyh} for future works.


\subsection{Recovering electromagnetism in a medium}\label{subsec:standardEM}

E\&M in a medium is one of the most studied theories ever. Understanding how it compares with the above construction thus seems unavoidable. All results can be written in three equivalent forms: form notation with Hodge dual and exterior derivatives, covariant notation with Lorentz indices and finally in terms of the good old electric and magnetic vector fields. Here are some of the constitutive equations
\begin{align}
&\text{Gauss law:} ~~~~~ d\star F= \star J \then \partial_{i} F^{i0} = \mu_0 J^0 \then \nabla \cdot \mathbf{E} = \frac{\rho}{\epsilon_0},\bigg. \\
&\text{Ampere law:} ~~~~~ d\star F= \star J \then  \partial_{\mu} F^{\mu i} = \mu_0 J^i \then -\frac{1}{c^2} \frac{\partial \mathbf{E}}{\partial t} + \nabla \times \mathbf{B} = \mu_0 \mathbf{J}, \bigg.\\
&\text{magnetic Gauss law:} ~~~~~ dF=0 \then \partial_{i}~\! ^\star\! F^{i0} = 0 \then \nabla \cdot \mathbf{B} = 0, \bigg.\\
&\text{Faraday induction:} ~~~~~ dF=0 \then  \partial_{\mu}~\! ^\star\! F^{\mu i} = 0 \then \frac{\partial \mathbf{B}}{\partial t} + \nabla \times \mathbf{E} = 0. \bigg.
\end{align}
The last two equations come from the Bianchi identity and do not involve any source. Consequently, they will not change in a material. 
Conversely, the first two equations come from the equations of motion, feature charges and current and change in the presence of a material.

A standard exercise consists in combining these linear equations in vacuum and solving them to find lightwaves propagating perpendicularly to $\mathbf{E}$ and $ \mathbf{B}$. In a material, the modifications of the the first two equations can be captured through the following manner. We introduce $ \mathbf{D}=\e \mathbf{E}$ with $ \e$ the electric permittivity and $ \mathbf{H}=\mathbf{B}/\mu$ with $ \mu$ the permeability.\footnote{This approach is actually only valid for the simplest ``linear'' material, where the response of the material's own electric and magnetic fields is proportional to $ \mathbf{E}$ and $ \mathbf{B}$. More generally, the presence of a material can induce non-linearities.} Then
\begin{align}\label{maxmedium}
\nabla \cdot \mathbf{D} &= \rho_{\text{free}} \qquad \text{and} \qquad \nabla \times \mathbf{H} - \frac{\partial \mathbf{D}}{\partial t} = \mathbf{J}_{\text{free}}
\end{align}
where $ \mathbf{J}_{free}$ and $ \rho_{free}$ are the currents and charges that can move around in the material (as opposed to being fixed somewhere like around an atom or a molecule). In the absence of free currents we can easily solve these equations again and find the speed of light in the material $ v^{2}=1/(\e \mu) $,
\begin{align}\label{eq:eomref}
\mu \e \ddot{\mathbf{E}}-\nabla^{2}\mathbf{E}=0\,.
\end{align}
The speed $ v$ is sometime related to the index of refraction $ n\equiv c/v$. The analysis of the dispersion relation obtained below \Eq{eq:disprelfin} revealed that the constant coefficient of the EFT parameter $\gamma_3$ directly relates to the speed of propagation in the medium through $\gamma_3 = - v^2$, that is $n = 1/\sqrt{-\gamma_3}$.

Higher derivatives tend to enrich the phenomenology while keeping the theory linear. One can simply work in frequency space in which all higher derivatives are collected into frequency and momentum dependence of the permittivity and permeability, $ \e=\e(\omega, k)$ and $ \mu=\mu(\omega, k)$. In Fourier space, both $ \e(\omega, k)$ and $ \mu(\omega, k)$ can be complex. It turns out that the imaginary part of $ \e(\omega, k)$ and $ \mu(\omega, k)$ leads to attenuation/dissipation, but also to a phase shift between the oscillations of $ \mathbf{E}$ and $ \mathbf{B}$.
From \Eq{eq:eomref}, the dispersion relation simply reads
\begin{align}
\mu(\omega,k) \e(\omega,k)\omega^{2} - k^{2}= 0\,.
\end{align}
Depending on the functions $ \e(\omega,k)$ and $ \mu(\omega,k)$, there can be many different phenomena, such as normal propagation or absorption, in a way that is different for different frequencies and/or different wavenumbers. 

We now see how to derive this phenomenology from the Schwinger-Keldysh formalism developed in the previous sections. To make contact with electromagnetism in a medium, it is useful to start from \eqref{OEMnog} rephrased in terms of $ F^{0i}=E^{i}$ and $ F^{ij}=\e^{ijl}B_{l}$. Adding the noise contributions discussed in \Sec{subsec:noise}, the constraint and equations of motion become
\begin{align}\label{eq:cEM}
\frac{\delta S_{\mathrm{eff}}}{\delta a^{0}}=0\quad &\then \quad \nabla. \mathbf{E}= j_{0} + \xi_0, \\
\frac{\delta S_{\mathrm{eff}}}{\delta a^{i}}=0\quad &\then \quad \gamma_{2}\mathbf{E} + \gamma_{3} \nabla \times \mathbf{B} - 2 \gamma_4 \mathbf{B} =\mathbf{j} + \mathbf{\xi}\,.
\end{align} 
Up to rescalings, we hence get equations similar to \eqref{maxmedium}, that is
\begin{align}
    \nabla \cdot \mathbf{D} &= \rho + \Xi \qquad \mathrm{and} \qquad \nabla \times \mathbf{H} - 2 \gamma_B \mathbf{H} + i \omega \mathbf{D}  = \mathbf{j} + \mathbf{\xi} \,.
\end{align}
To reach these expressions, we identified the permeability $\mu = 1/\gamma_3$, the permittivity $\epsilon = \gamma_2/(i\omega)$ and the birefringence index $\gamma_B = \gamma_4/\gamma_3$. We also used
the redefinitions $\rho = \epsilon j_0 $ and $\Xi = \epsilon \xi_0$. 
Apart from birefringence, the main difference with \Eq{maxmedium} comes from the stochastic contributions $\Xi$ and $\mathbf{\xi}$ which model the presence of random impurities in the medium through which light propagates.  Explicitly, \Eq{maxmedium} is recovered by setting $\gamma_B = \Xi = \mathbf{\xi} = 0$. 

In summary, the standard textbook treatment of electromagnetism in a medium is easily recovered from the open EFT construction. The effective functional $S_{\mathrm{eff}}$ generates modified Gauss and Ampère laws, accounting for the propagation of light in a dispersive ($\gamma_3$), dissipative ($\gamma_2$), anisotropic ($\gamma_4$) and random ($\xi_\mu$) medium.




\section{Conclusion}\label{sec:conclu}

In this article, we develop a bottom-up Open EFT for the propagation of light in a medium experiencing dissipation and noise. This work contributes to establish the first foundations to a systematic study of open quantum systems that involve gauge fields in the Schwinger-Keldysh formalism. Our ultimate goal is the description of open gravity, a theory of dynamical gravity in the presence of an unknown environment. Open E$\&$M is a first step in this direction, providing an explicit and pedagogical example of such construction. Complementing formal approaches in non-equilibrium EFTs and dissipative hydrodynamics \cite{Crossley:2015evo, Glorioso:2016gsa, Liu:2018kfw, Hongo:2018ant, Chen-Lin:2018kfl, Hongo:2019qhi, Armas:2020mpr, Landry:2021kko, Vardhan:2024qdi, Abbasi:2024pwz, Hongo:2024brb, Liu:2024tqe, Baggioli:2024zfq} — such as the in-in coset construction \cite{Akyuz:2023lsm} and Schwinger-Keldysh BRST quantization \cite{Haehl:2016pec} — this work sheds light on gauge symmetries in the closed-path time contour for open systems. 

\paragraph{Summary.}

We start by reviewing the main features of the Schwinger-Keldysh formalism. In \Sec{sec:SK}, we first outline in \Eq{eq:NEQ} the constraints any open effective functional has to follow. We then introduce the Keldysh basis in \Eq{eq:RAbasis} for the doubling of the degrees of freedom. The boundary conditions for the path integral are carefully discussed in \Sec{subsec:BCs}. We make a connection with another description of stochastic systems, namely the Langevin equation, which is derived via a saddle point approximation. The Hubbard-Stratonovich trick presented in \Eq{eq:HStrick} provides an interpretation to some of the non-unitary terms as sourcing a stochastic variable for the equations of motion. We then discuss the dispersion relation of a toy model with a single dissipative scalar. At last, we close \Sec{sec:SK} explaining how to relate the correlation functions of the variables of the Keldysh basis with the expectation values of quantum mechanical operators.

In \Sec{sec:scalars}, we carry out the analysis of the number of degrees of freedom of a dissipative free theory specified by its Hessian matrix $M$ that defines its equations of motion. We start from a generic $M$ and establish the conditions under which the time evolution remains unitary. When unitarity is broken, the dissipative character of the system may arise in two ways: via non-reciprocity (which preserves time-reversal symmetry), and via time-translation symmetry breaking. We give toy models for both scenarios, see \Eqs{eq:Mdeom} and \eqref{eq:Tbreaking}. We then present in \Eq{mex} an algorithm to derive the number of degrees of freedom from a generic diagonalizable $M$ matrix, and discuss its extension to the fine-tuned case of non-diagonalizable matrices in \Eq{dofgeneral}. We apply these formulas to three systems with constrained degrees of freedom, illustrating the well-known cases of Proca theory and the scalar sector of gravity.

\Sec{sec:OEM} constitutes the main result of this work. We first construct the most generic linear functional, using retarded gauge invariance \eqref{eq:gaugetrans} as a guiding principle. Studying invariance under this gauge transform, we uncover a second — advanced — gauge invariance, spanned by a different eigenvector denoted $v^{\mu}$, as summarized in \Eq{eq:advgauge}. The advanced gauge invariance generates a constraint on the noise presented in \Eq{eq:noiseconstraint}, a generalization of the conservation equation for currents to dissipative systems. At zero dissipation where we recover the unitary results. More surprisingly we discover that dissipation does not break the symmetry to the diagonal subgroup, but leaves it modified and unbroken. In particular, we can still gauge transform along each contour of the path integral and, in this way, we can fix the gauge in both the advanced and the retarded components. 

For the linear dynamics, the gauge fixing procedure reduces the matrix $M$ from a $4\times4$ matrix that is not invertible to a $3\times3$ invertible matrix. The noise constraint \eqref{eq:noiseconstraint} projects out the physical fluctuations on the direction orthogonal to the advanced gauge orbit. In practice, it explains why it is equivalent to either first gauge fix then add the noises, or to first add the noises then gauge fix. This approach allows us to solve the equation of motion by just inverting this new $3\times3$ matrix. The resulting degrees of freedom are indeed a constrained mode and two propagating degrees of freedom, as shown in \Eq{eq:disprelfull}. Studying the dispersion relations for the propagating degrees of freedom, we become able to physically interpret the Wilson coefficients of \eqref{OEMnog} and to place bounds on them. In particular, we identify the birefringence and dissipation terms, the first affecting each polarization in a different way and the latter generating a damping of the signal. 

In \Sec{subsec:noise}, we derive the noise constraint from a path integral approach, reconciling the existence of a modified conservation law for Open E\&M with the non-equilibrium constraints. In \Sec{subsec:SantiEM}, we present an alternative derivation of the effective functional given in \Eq{OEMnog} from gauge invariance, Lorentz invariance and unitarity considerations. In \Sec{subsec:covariant gauges}, we study the covariant gauge formulation of Open E\&M. We start by reviewing the usual gauge fixing $\zeta$-term in Maxwell theory and how gauge invariance is encoded by the $\zeta$-independence of the time-ordered correlation function of the field strength. We then analyze the degrees of freedom for Open E\&M with the $\zeta$-term, finding the expected result — 1 ghost, 1 constraint and 2 propagating degrees of freedom. Paving the way for future investigation of interactions, we derive the retarded and Keldysh propagators in \Eqs{eq:retfull} and \eqref{eq:keldyshinv} respectively, and check their gauge invariance. In principle, inclusion of the $\zeta$-term introduces a ghost mode that also removes the noise constraint. However, we show that the Gupta-Bleuler condition \eqref{eq:newGPcondition}, which removes the ghost in the unitary theory, also generates the noise constraint we derived in \Sec{subsec:noise}. 

At last, we discuss the inclusion of topological terms, first recovering the usual operator from unitary theories in \Eq{eq:thetauni} and then uncovering a new topological term in \Eq{eq:thethanoise}, unique to non-unitary theories. We close \Sec{sec:OEM} by discussing how to recover the textbook description of E\&M in a dielectric medium.


\paragraph{Discussion.} The results presented in this work rely on a series of assumptions. First, we have focused on time independent dielectric media, such that there are only two propagating degrees of freedom. This would change if the media was a conductor, a superconductor or a plasma. Second, since the theory studied in this paper being a free theory, we have truncated the open effective functional to quadratic order and focused on the counting of degrees of freedom. The constraints imposed by gauge invariance holds beyond the free theory, so they must survive the inclusion of interactions. Future work would study interactions consistent with the advanced gauge transformation and the inclusion of loops effects. This would generate perturbativity constraints on the couplings and open a window onto the study of anomaly cancellation in the presence of dissipation and diffusion. 

The main objective of this program is to develop a dissipative theory of open gravity. In that respect, our most interesting result here is the noise constraint \eqref{eq:noiseconstraint}. This constraint is instrumental in propagating physical noise-sourced degrees of freedom. Furthermore, a correct counting of degrees of freedom in open systems as the one proposed in \Sec{ssec:Jordan} is fundamental for the study of open gravity. Indeed, the scalar potentials $\phi$ and $\psi$ should remain constrained while we open the gravitational system, albeit modified to account for noises and dissipation, just as the vector constraints we found in this work. Furthermore, a generic theory of open gravity would \textit{a priori} have $10$ different stochastic noises. Only through a constraint similar to \Eq{eq:noiseconstraint} the number of sourced degrees of freedom can match those in the gravitational sector. At last, a theory of gravity that treats the metric fluctuations as an open system will also include background quantities, such as energy densities or a cosmological constant. These quantities have to obey a background equation given by the tadpole cancellation of the open effective functional. This tadpole has not been discussed in this work yet seems to play an instrumental role in a theory of open gravity. Interestingly, fluctuations of the background quantities are not part of the system's degrees of freedom but rather enter the generalization of the noise constraint \eqref{eq:noiseconstraint}, very much like the charged current $j_{\mu}$ does in this work. 

A theory of open gravity has many different cosmological applications, for example the propagation of gravitational waves in an interacting dark energy scenario or the coupling to the inflaton field in a dissipative inflation scenario. These scenarios will lead to an extension of the work carried out in \cite{Salcedo:2024smn} beyond the assumed decoupling limit. 

\paragraph{Phenomenology directions.} This work is the first step towards a proper understanding of gauge symmetries in open quantum systems using the Schwinger-Keldysh formalism. A natural generalization consists in extending this approach to Yang-Mills theories. There, the non-Abelian character of the symmetry group makes the retarded gauge transformation more complicated, 
\begin{equation}
    A\rightarrow U A U^{\dagger} - U^{\dagger} dU \quad,\quad a\rightarrow U a U^{\dagger}.
\end{equation}
where $d$ is the exterior derivative. This transformation makes the construction of an open effective functional more involved. Yet, this extension opens the possibility to study gauge anomalies in the presence of dissipation. One of the most important constraints one can impose on a gauge theory is the absence of gauge anomalies. This can be a crucial consistency check for our theories. A correct description of this phenomenon is instrumental to study realistic open gauge field theories like QCD at finite density.

Another extension of this work consists in a field theoretical study of plasma oscillations, where a description based on Open scalar QED and Open scalar QCD will be crucial. However, the study of open gauge theories is not limited to plasma physics. The photons in the Cosmic Microwave Background (CMB)  and any light traveling through a cosmological medium are very likely to be described by an open Abelian gauge theory. It will be interesting to see if we can combine the field theoretic description of this work with the current understanding of CMB photons in terms of geodesic equations implemented in a Boltzmann code. Furthermore, photons being conformally coupled, most of the developments carried out in this work should be extendable to FLRW cosmologies, with an emphasis on the inflationary phase and the late universe. This extension will contribute to the recent investigation of the role of open dynamics and out-of-equilibrium processes in cosmology \cite{Lombardo:2005iz, Burgess:2006jn, Anastopoulos:2013zya, Fukuma:2013uxa, Burgess:2014eoa, Burgess:2015ajz, Boyanovsky:2015tba, Boyanovsky:2015jen, Boyanovsky:2015xoa, Liu:2016aaf, Nelson:2016kjm, Hollowood:2017bil, Shandera:2017qkg, Boyanovsky:2018fxl, Boyanovsky:2018soy, Martin:2018zbe, Bohra:2019wxu, Akhtar:2019qdn, Kaplanek:2019dqu, Brahma:2020zpk, Burrage:2018pyg, Burrage:2019szw, Kaplanek:2020iay, Rai:2020edx, Geng:2020qvw, Jana:2020vyx, Geng:2021hlu, Burgess:2021luo, Kaplanek:2021fnl, Brahma:2021mng, Zarei:2021dpb, Banerjee:2021lqu, Brahma:2022yxu, Kaplanek:2022xrr, Kaplanek:2022opa, Colas:2022hlq, Sou:2022nsd, Colas:2022kfu, DaddiHammou:2022itk, Burgess:2022nwu, Burgess:2022rdo, Cao:2022kjn, Prudhoe:2022pte, Kading:2022jjl, Kading:2022hhc, Danielson:2022sga, Danielson:2022tdw, Gong:2023kpe, Colas:2023wxa, Brahma:2023hki, Sharifian:2023jem, Alicki:2023tfz, Alicki:2023rfv, Ning:2023ybc, Geng:2023ynk, Geng:2023zhq, Kading:2023mdk, Sou:2024tjv, Burgess:2024eng, Bowen:2024emo, Tinwala:2024wod, Keefe:2024cia, Hsiang:2024qou, Proukakis:2024pua, Colas:2024lse, Salcedo:2024smn, Liang:2024shg, Brahma:2024yor, Danielson:2024yru, Biggs:2024dgp, deKruijf:2024ufs, Ota:2024yws, Brahma:2024ycc, Ota:2024mps, Kading:2024jqe, Green:2024cmx, Ivo:2024ill}.

\paragraph{Formal extensions.} The formalism developed in this work relies on treating the gauge field $A^{\mu}$ as a Lorentz 4-vector. However, this is not the most condensed formalism to describe gauge theories and it is not the most suitable one for non-Abelian gauge theories. An extension to the latter will greatly benefit from finding a differential geometry approach to Open E\&M. In particular, such an approach could provide a geometric interpretation to the advanced gauge transformation as a deformed fiber bundle. 

The existence of noise constraint is at the center of the correct matching of degrees of freedom between noise variables and the retarded field $A^{\mu}$. In \Sec{subsec:covariant gauges}, we gave a first derivation of this constrain as a quantization condition for the covariant gauges. However, this is only a minimal extension and does not leverage the Dirac formalism derived for constrained systems. Similarly, in \Sec{ssec:Jordan}, we provided an algorithm to count the number of degrees of freedom giving a matrix $M$ that describes the equations of motion for the system. Still, this is not the most general way to count the degrees of freedom of the system, as Hamiltonian analysis provides a more rigorous and robust way to carry out this counting. A phase-space extension of the current formalism would then be valuable.

Open quantum systems may obey further constraints if they reach a state of equilibrium with their environment. Future work may leverage the use of detailed balance and fluctuation-dissipation relations in open EFTs \cite{Ota:2024mps}. In particular, non-equilibrium effective field theories have been studied in detailed for dissipative hydrodynamics \cite{Vardhan:2024qdi, Liu:2024tqe, Haehl:2016pec, Hongo:2024brb}. In this case, the object of study can also be a vector field. Therefore, it is instrumental to properly understand the overlap between our work and developments in this field. Similarly, future extensions would benefit from the expertise of the photonics community, connecting Schwinger-Keldysh path integral and master equation treatments \cite{breuerTheoryOpenQuantum2002, breuerTimelocalMasterEquations2002, Colas:2022hlq, Khatiwada:2024gql, Carney:2024izr, Kashiwagi:2024fuy}.

At last, the stability and causality constraints on the Wilson coefficients discussed in \Sec{subsec:disprel} rely on very general principles. There has been recent works using a positivity bounds approach to electromagnetism in a medium \cite{Creminelli:2022onn, Creminelli:2024lhd}. This presents an opportunity to exploit the interplay between open quantum systems and scattering amplitudes to find a window into constraints beyond \eqref{eq:NEQ} and those relaying on detailed-balance or fluctuation-dissipation relations. Open quantum systems are described in terms of a density matrix $\hat{\rho}$ that verifies various set of constraints. One of them is a bound on the purity, $0<\text{Tr}[\hat{\rho}^{2}]\leq1$. The computation of this quantity in cosmological settings has received a lot of attention in recent years \cite{Pelliconi:2023ojb, Pelliconi:2024aqw, Maldacena:2012xp, Colas:2022kfu, Colas:2024xjy, Burgess:2024eng, Burgess:2024heo, Colas:2024ysu} and has been used to constrain the Wilson coefficients of a variety of effective field theories \cite{Pueyo:2024twm}. Future works way bound open EFT coefficients from entropy measures. 


\subsection*{Acknowledgements} We thank Tarek Anous, Michael Cates, Sean Hartnoll, Maria Mylova, Toshifumi Noumi, Riccardo Penco, David Tong, Gonzalo Villa and Manuel Loparco for the insightful discussions. This work has been supported by STFC consolidated grant ST/X001113/1, ST/T000694/1, ST/X000664/1 and EP/V048422/1. S.A.S. is supported by a Harding Distinguished Postgraduate Scholarship. 
 

\appendix



%
\bibliographystyle{JHEP}
\bibliography{Biblio}
%


\end{document}